# Label-free incoherent super-resolution optical microscopy

Nikhil Jayakumar**[1], Luis E. Villegas-Hernández[1], Weisong Zhao[2], Hong Mao[1], Firehun T Dullo[3], Jean-Claude Tinguely[1], Krizia Sagini[4,5], Alicia Llorente[4,5,6], Balpreet Singh Ahluwalia*[1,7]

[1]*Department of Physics and Technology, UiT The Arctic University of Norway, Tromsø, 9037, Norway*
[2]*Innovation Photonics and Imaging Center, School of Instrumentation Science and Engineering, Harbin Institute of Technology, Harbin, China*
[3]*Department of Microsystems and Nanotechnology, SINTEF Digital, Gaustadalleen 23C, 0373 Oslo, Norway*
[4] *Department of Molecular Cell Biology, Institute for Cancer Research, Oslo University Hospital, The Norwegian Radium Hospital, 0379 Oslo, Norway*
[5]*Centre for Cancer Cell Reprogramming, Faculty of Medicine, University of Oslo, Montebello, 0379 Oslo, Norway*
[6]*Department for Mechanical, Electronics and Chemical Engineering, Oslo Metropolitan University, Oslo, Norway*
[7]*Department of Clinical Science, Intervention and Technology, Karolinska Institute, Sweden*

*Balpreet.singh.ahluwalia@uit.no, **nik.jay.hil@gmail.com

**Abstract**: The photo-kinetics of fluorescent molecules have enabled the circumvention of far-field optical diffraction-limit. Despite its enormous potential, the necessity to label the sample may adversely influence the delicate biology under investigation. Thus, continued development efforts are needed to surpass the far-field label-free diffraction barrier. The coherence of the detected light in label-free mode hinders the application of existing super-resolution methods based on incoherent fluorescence imaging. In this article, we present the physics and propose a methodology to circumvent this challenge by exploiting the photoluminescence of silicon nitride waveguides for near-field illumination of unlabeled samples. The technique is abbreviated EPSLON, Evanescently decaying Photoluminescence Scattering enables Label-free Optical Nanoscopy. We demonstrate that such an illumination has properties that mimic the photo-kinetics of nano-sized fluorescent molecules. This allows for developing a label-free incoherent system that is linear in intensity, and stable with time thereby permitting the application of techniques like structured illumination microscopy (SIM) and intensity-fluctuation-based optical nanoscopy (IFON) in label-free mode to circumvent the diffraction limit. We experimentally demonstrate label-free super-resolution imaging of nanobeads (polystyrene and gold), extra-cellular vesicles, rat kidney sections, human kidney sections and human placenta tissue. In this proof-of-concept work, two-point resolution of $\sim$180 nm on nanobeads, $\sim$133 $nm$ mean Fourier Ring Correlation (FRC) resolution ($\sim$1.92 times resolution gain) on rat kidney tissue sections, $\sim$129 $nm$ mean FRC resolution on human kidney tissue sections ($\sim$2.8 times resolution gain) and $\sim$176 nm mean FRC resolution on human placenta tissue ($\sim$2.8 times resolution gain) is demonstrated. We believe EPSLON is a step forward within the field of incoherent far-field label-free super-resolution microscopy that holds a key to investigating delicate biological systems in their natural state without the need for exogenous labels.

## Introduction

The ability of light beams to interfere is quantified by their degree of coherence. Light beams originating from within the coherence volumes can only overlap and generate a sustained interference pattern [1, 2]. In fluorescence microscopy, the transversal coherence lengths are typically on the order of a few nanometers. This is because the fluorescent molecules, a few nanometers in size, emit independently and stochastically. It leads to a linear mapping between the sample plane fluorophore concentration and image plane intensity. This may be utilized to circumvent the far-field diffraction-limit, as in structured illumination microscopy [3, 4] or fluorescence-based IFON algorithms [5-10]. However, the absence of such exogenous molecules in label-free microscopy restricts the far-field transversal coherence lengths to a few hundreds of nanometers [11, 12]. This hinders the application of fluorescence-based super-resolution algorithms in the label-free regime for generating reliable super-resolved images [13]. Another hinderance in label-free microscopy is the lack of selectivity and specificity that results in strong scattering and multiple scattering from the entire sample. To alleviate these challenges of scattering, a near-field illumination via nano-sized light sources with stochastic photo-kinetics and sufficient quantum yield will be beneficial. Therefore, through this article, we provide the concepts and a key to unlock the challenge of generating far-field label-free super-resolved optical images using fluorescence-based super-resolution algorithms: photoluminescence (PL) of silicon nitride ($Si_3N_4$) [14, 15] waveguide functions as exogenous nano-sized illumination sources with stochastic photo-kinetics. In addition, the photonic-chip helps in engineering the illumination to induce fluctuations in intensity via multi-mode interference (MMI) speckle-like patterns [16-18] or via well-defined interference fringes that permits the application of fluorescence-based IFON algorithms [19, 20] or SIM respectively to enhance the resolution.

Poor-contrast and diffraction-limited resolution are major impediments to the development of label-free optical microscopy. To mitigate the issue of poor contrast, various approaches have emerged: phase contrast microscopy [21], differential interference contrast [22], Hoffman modulation [23], interferometric scattering microscopy [24], quantitative phase microscopy [25], holographic non-interferometric techniques [26], Fourier Ptychography [27], rotating coherent scattering microscopy [28], manipulating the coherence of light sources used for illumination [29], ultraviolet microscopy [30], optical waveguides [31, 32], among others. However, circumventing the diffraction-limit in label-free regime is still challenging in life sciences, as opposed to fluorescence microscopy [33]. This could be attributed to the ease of utilizing/manipulating the photo-kinetics of nano-sized fluorescent molecules to gain information beyond the diffraction-limit.

The different approaches developed for label-free super-resolution microscopy, albeit with their respective experimental challenges especially for life sciences applications, includes near-field scanning optical microscopy [34], super-lens [35], micro-sphere assisted super-resolution imaging [36], high-index liquid immersed microsphere assisted super-resolution [37], optical super-oscillation techniques [38], non-linear imaging systems [39], multiplexing information in polarization, wavelength or time [40], etc.

For the far field high-resolution label-free optical microscopy techniques, different concepts have been developed [41]. A short overview of different approaches is detailed in the Supplementary section 1. Broadly, these methods employ the concept of synthetic aperture or spatial frequency shifting for coherently scattering specimens using free-space optics [27, 28, 42-44] or chip-based solutions [31, 45-47], or application of fluorescence-based super-resolution algorithms to coherently scattering specimens [48]. Some of these techniques achieve sub-100 nm resolution, but the best achievable theoretical resolution is given by Abbe's diffraction-limit, $\frac{\lambda_{det}}{NA_{ill}+NA_{det}}$, where $\lambda_{det}$ is the wavelength of the detected light and $NA_{ill/det}$ is the numerical aperture of the illumination and detection light paths, respectively. In addition, some methods indeed achieve label-free super-resolution by utilizing the intrinsic auto-fluorescence of biological specimens in tandem with super-resolution fluorescence-based algorithms [49, 50]. In this article, we propose the use of photoluminescence of $Si_3N_4$ waveguides to solve the abovementioned challenges associated with the far-field label-free super-resolution optical microscopy. The concept permits the application of fluorescence-based super-resolution algorithms on unlabeled samples, generating high-contrast label-free super-resolved images, without photo-toxicity and photobleaching plaguing the imaging process. Our work helps synthesize a label-free incoherent imaging system and is termed Evanescently decaying Photoluminescence Scattering enables Label-free Optical Nanoscopy (EPSLON), which builds and extends the concepts outlined by Ruh et.al. [25], Wicker and Heinztmann [13], and previous work based on photonic-chip microscopy [29]. These concepts of EPSLON which enable circumventing the label-free far-field diffraction-limit are explained and experimentally demonstrated in Fig. 1 and Fig. 2. We describe how PL from $Si_3N_4$ waveguide is a solution to these challenges and validate our concepts experimentally via high-contrast label-free super-resolved images of polystyrene nanobeads, gold nanoparticles, weakly scattering specimen like extra-cellular vesicles, human placenta tissue and rat kidney sections.

## Conceptual Framework and Results

*Problem statement*

Here we describe why fluorescence-based super-resolution algorithms when applied to coherently scattering samples do not yield any resolution gain beyond the diffraction-limit [13]. In label-free mode, when two particles are illuminated by a monochromatic plane wave, the intensity registered by the camera is $I(\vec{r}) = \langle \left| \left( E_1(\vec{r_1}, t) + E_2(\vec{r_2}, t) \right) \otimes h(\vec{r}) \right|^2 \rangle$, where $\langle\ \rangle$ represents time averaging by the detector, $h(\vec{r})$ is the coherent point spread function of the imaging system , $\otimes$ represents the convolution operation and $E_{1,2}$ are the scattered scalar electric fields which are linked to applied electric fields via polarizability of the two particles [51]. Due to statistical similarity or coherence between the overlapping scattered fields, the intensity registered by the camera is non-linearly related to the particle concentration and is a function of $\Delta\varphi = \vec{k}.\vec{r_2} - \vec{k}.\vec{r_1}$. It implies that the image generated by the camera varies with either a change in the illumination angle $\vec{k}$, or with the relative positions of the particles $\vec{r_2} - \vec{r_1}$.

Next, to illustrate the image formation process in fluorescence microscopy we replace the phase-objects with fluorescent molecules. Analogous to the strengths of the scattered fields, $|a_1(\vec{r_1})|^2$ and $|a_2(\vec{r_2})|^2$, are the brightness of the molecules that typically depends on the illumination intensity at the location of the molecule. The molecules can also be assumed to emit independently [52] and stochastically typically on the order of nanoseconds [53]. The following properties of these molecules can be utilized in fluorescence microscopy by collecting only the Stoke shifted light emitted by the molecules:

(i) Stochastic emission between the molecules causes the phase difference between the emitted fields to be a function of time, $\Delta\varphi(t)$. It implies that the molecule emissions are incoherent with respect to one another, or, in other words, we can say that the transversal coherence length is determined by the size of an individual molecule. This gives rise to similar images for different illumination angles of the incident plane wave. It is this property that allows the usage of structured light in SIM [13, 54] or the intrinsic photo-kinetics of the molecules in IFON algorithms to enhance the resolution [29],

(ii) Excited lifetime on the order of nanoseconds of these molecules helps mitigate the speckle-noise and

(iii) Molecular specificity offered by these fluorescent molecules enables multi-color imaging of different cell organelles. It can be concluded that the suppression of speckle noise and molecular specificity offered by the molecules enables high-contrast imaging and the linear relationship between molecular concentration and image plane intensity helps in enhancing the resolution via fluorescence-based super-resolution algorithms.

Hence, to improve the label-free resolution via fluorescence-based algorithms, we need to ensure that there exists no statistical similarity between the scattered fields originating from different locations [51]. This calls for the mutual intensity to be δ-function correlated, i.e.,

$$J(\vec{r_1},\vec{r_2}) = \langle E_T(\vec{r_1},t)E^*{}_T(\vec{r_2},t)\rangle = KI_T(\vec{r_1})\,\delta(\vec{r_1}-\vec{r_2}) \quad (1)$$

where $J(\vec{r_1},\vec{r_2})$ is the mutual intensity and determines the spatial correlation of the fields, $E_T(\vec{r},t) = E_1(\vec{r},t) + E_2(\vec{r},t)$ is the total field reaching the camera, $K$ is a real constant and $I_T$ is the image generated by the camera. This will ensure an incoherent imaging system. Eqn. (1) can be assumed to be satisfied in fluorescence microscopy because the transversal spatial coherence length is determined by the size of the fluorescent molecules and the image generated by the camera indicates the spatial locations of the fluorescent molecules.

Thus, to circumvent the label-free diffraction-limit using fluorescence-based super-resolution algorithms, we need to develop a light source whose electric field has δ-function spatial correlations, and then acquire an image stack exhibiting intensity-fluctuations to apply SIM or intensity-fluctuation based algorithms. This can be realized experimentally via the EPSLON configuration. Fig. 1(a-c) compares the conventional imaging configurations and their corresponding image plane intensity distribution with EPSLON, Fig. 1(d). EPSLON satisfies Eqn. 1 and is experimentally demonstrated in Fig. 1e-1j. In Fig. 1e-1f, schematic diagrams of waveguide-based label-free coherent and incoherent imaging systems EPSLON are shown. The coherent and corresponding EPSLON images are compared in Fig. 1g, where speckle suppression due to loss in phase information in the scattered light is clearly evidenced in the EPSLON configuration image. This loss in phase information also implies that identical images must be generated for arbitrary illumination angles in EPSLON configuration, as opposed to label-free coherent imaging. This is demonstrated experimentally in Fig. 1h-1j where the coherent and its corresponding incoherent EPSLON images are provided.

*EPSLON: a solution for high-contrast far-field label-free super-resolution microscopy*

To employ a light source whose electric field has δ-function correlations, we resort to the high-index contrast ($\Delta$n ≈ 2) $Si_3N_4$ optical waveguide deposited using plasma enhanced chemical vapor deposition (PECVD) scheme. Waveguide fabrication and properties of the guided modes and its spatial frequency extend is provided in Supplementary section 2 and in previous works [55, 56]. The propagation loss in these waveguides as a function of wavelength is determined and given in Table 1 in the supplementary section 2. The choice of $Si_3N_4$ over other high-index contrast optical waveguides, such as tantalum pentoxide, $Ta_2O_5$, or titanium dioxide $TiO_2$, is attributed to the room-temperature visible PL generated inside the core during the transfer of optical power along its length.

Determining the origin and lifetime of this emission is not within the scope of this work. The origin and photophysical properties of this PL is a widely researched area [57-59]. It is found to be dependent on the waveguide fabrication scheme employed and could be attributed to the intrinsic fluorescence of the material [14]. The PL emission spectrum is broad [15] and the lifetime of these states is found to vary on the order of a few nanoseconds to a few hundred microseconds depending on the origin of the PL [16, 60-61]. Such an emission could be visualized as a very large number of fluorescent molecules embedded in a material and emitting stochastically. Hence, if the PL light is used for near-field illumination of samples, then Eqn. (1) will be satisfied for the incoherently scattered fields. This helps in synthesizing a label-free incoherent system, Fig. 1.

The next problem to tackle is that of generating an image stack with intensity-fluctuations for the fluorescence-based algorithms. Structuring the illumination beam, manipulating the photophysical properties of the fluorescence molecules are some of the ways typically employed for generating image stacks with intensity-fluctuations. In EPSLON, this problem is resolved by resorting to $Si_3N_4$ waveguides of the following types: (1) Straight waveguides with strip geometry and large widths that support a large number of the guided modes, generating MMI patterns (Fig. 2a) [32], (2) Four-arm junction multi-moded strip waveguides for speckle illumination from different azimuthal angles (Fig. 2b) [32] and, (3) a single moded SIM chip with rib geometry and phase modulation for one-dimensional structured illumination (Fig. 2c) [62] and four-arm junction multi-moded strip waveguide for two-dimensional structured illumination. Simulation analysis and experimental results are presented in the Supplementary Section 3 to validate how multi-moded illumination patterns when used in tandem with IFON help gain resolution. Our results in Supplementary Fig. 3-6, show that the different azimuthal illumination frequencies in four-arm junction multi-moded waveguide, help achieve low correlation between different scatterers and thereby aid, IFON techniques like SOFI [5] and SACD [9] that exploit intensity-correlations between different emitters to generate super-resolved images.

*Image formation process in EPSLON*

Laser is coupled into a $Si_3N_4$ waveguide via a microscope objective $MO_1$, (Supplementary Fig. 7). The coupled optical power gets distributed among the multiple coherent guided modes, which correspond to the eigen vectors of the guiding structure. As the modes guide power along the length of the waveguide, they also induce a broadband PL along the length as shown in Fig. 2d-2e. The $Si_3N_4$ waveguide employed in this work demonstrated PL in all the commonly used wavelengths in bio-imaging, 488 nm, 561 nm, and 647/660 nm (Fig. 2e). This PL of $Si_3N_4$ waveguide does not exhibit any bleaching effect over long periods of times (Fig. 2f), as opposed to the autofluorescence in polymer waveguides [63] and varies linearly with the excitation power (Fig. 2g).

To explain the origin of fluctuations in intensity, we consider the case of a multi-moded straight waveguide. Fluctuations over time, $I^m_{core}(\vec{r},t)$, can be induced by oscillating $MO_1$ using a piezo-stage across the input facet of the waveguide (Supplementary Fig. 7) which excites different sets of modes $\psi_m(\vec{r},t)$ with relative amplitudes $0 \leq a_m(t) \leq 1$. The instantaneous PL intensity at each location in the core is dependent on the coherent superposition of the modes and can be represented mathematically as $I^m_{core}(\vec{r},t) = \eta(\lambda)|\sum_m a_m(t)\psi_m(\vec{r},t)|^2$, where $\eta(\lambda)$ is assumed to be a constant across the material for a specific wavelength. $\psi_m(\vec{r},t) = E_m(x,y)e^{(i ß_m z - i\omega t)}$ corresponds to the scalar representation of the m$^{th}$ guided mode with fixed transversal profile $E_m(x,y)$ and propagation constant $ß_m$. As the PL emission occurs inside a high-index core, a part of the PL light gets confined to the core due to total-internal reflection at the core-cladding interface and the remaining part gets transmitted into the far-field, which is visible as an omnipresent background or noise. This ratio is quantified experimentally and found to be about 0.01, Supplementary Fig. 8a. What this implies is that the scattered light off the sample in EPSLON configuration will be stronger than the background. This can also be understood from Supplementary Fig. 8b, where tissue scattering is predominantly due to confined light in the waveguide.

Now the presence of any index perturbation at the core-cladding interface scatters this evanescently decaying PL light into the far-field (Fig. 1d) and hence the technique is abbreviated EPSLON. That is, both the coherent as well as the Stoke shifted incoherent PL light gets scattered into the far-field. By invoking a first order Born approximation for evanescent field excitation of biological specimens $S(\vec{r}) = \iint_S \alpha(\vec{r}_k)\delta(\vec{r} - \vec{r}_k)\, d\vec{r}_k$ with α being the polarizability, it is seen that these scattered fields contain the information of the sample. Only these scattered fields are collected by the microscope objective $MO_2$ and relayed onto the camera via tube lens because of the decoupled illumination and detection scheme offered by waveguides (Supplementary Fig. 7). Through the usage of appropriate bandpass filters, the coherently scattered light is filtered out and only the incoherent light gets detected. The oscillation of $MO_1$ is synchronized with the detector in such a way that an image is acquired at each excitation point of the waveguide. By invoking Eqn. (1) and neglecting noise, an EPSLON image (Fig. 1d) at the camera plane in general may be described mathematically as

$$I^m(\vec{r}) = \eta|S(\vec{r})\sum_m a_m(t)\psi_m(\vec{r},t)|^2 \otimes |h(\vec{r})|^2 \tag{2}$$

The abovementioned concepts and experimental results can be summarized into the following: (i) Speckle noise is mitigated in EPSLON images as opposed to label-free waveguide-based coherent images due to stochastic fluctuations between the scattered fields, Fig. 1g. (ii) Stochastic fluctuations implies that phase relationships between the scattered fields are lost in EPSLON images as opposed to label-free coherent images. This also leads to identical images for different illumination angles, Fig. 1(h-j) and (iii) Intensity-fluctuations are induced over time due to time dependence of $a_m(t)$ and $\psi_m(\vec{r},t)$. This is evident in the line plots in Fig. 1j and in the MMI patterns shown as insets in Fig. 2a-2c. This is also validated using simulations in Supplementary section 5, Fig. 9. Thus, EPSLON when used in tandem with fluorescence-based super-resolution algorithms helps develop a high-contrast label-free super-resolution imaging system.

*Applications of EPSLON*

The potential of EPSLON is first demonstrated on 195 nm polystyrene beads. For brevity, the diffraction-limited label-free image and its corresponding reconstructed super-resolved image in the following sections are termed DL and EPSLON. 2D label-free SIM is demonstrated using four-arm junction waveguide as shown in Fig. 3a and in Supplementary Fig. 10. The use of such a waveguide geometry helps in introducing additional illumination frequencies (Supplementary Fig. 2). 50 images each are acquired using a 0.75 NA detection objective (Fig. 3a) and a 0.9 NA detection objective (Supplementary Fig. 10). The corresponding pixel sizes at the sample plane for 0.75 NA and 0.9 NA detection objectives are 325 nm and 108 nm respectively. The EPSLON images are then generated using BlindSIM reconstruction algorithm [64]. The EPSLON images of 0.75 NA are validated using a 1.2 NA detection objective which serves as the ground truth optical image. Three insets in the DL image labelled '1', '2' and '3' are blown up and shown. The line plots correspond to intensity variations along the red and green arrows in the DL and EPSLON images respectively. It is seen that the use of such a waveguide geometry helps to resolve beads oriented along different azimuthal directions. EPSLON resolves beads separated by 567 nm, 325 nm and 517 nm in insets '1', '2' and '3' respectively as shown in Fig. 3a. The EPSLON and ground truth images are in good agreement. Then a region within this same field-of-view is then imaged with a 0.9 NA detection objective and the corresponding EPSLON image is shown in Fig. 10 in the supplementary section. Next, in the supplementary Fig. 11, we demonstrate label-free 1-D SIM in EPSLON configuration using three phase-shifted frames as input to FairSIM [65]. The nanobeads are deposited directly on the core of SIM structure shown in Fig. 2c. The angle between the interfering waveguides used in the experiment is 60º. This will create an interference pattern with a fringe period $f = \frac{\lambda_{ex}}{2n_f \sin\frac{\theta}{2}}$ where $\lambda_{ex}$ is the excitation wavelength, $n_f \approx 1.7$ is the refractive index of the guided mode and $\theta$ is the angle between the interfering waveguides. For better visualization purposes, the interference fringe pattern shown in Fig. 2c is generated by waveguides interfering at an angle $\theta = 20°$. In supplementary section 8 (supplementary Fig. 12 – Fig. 15), the dependence of speckle size and resolution on $\lambda_{ex}$, $\theta$ and $\lambda_{det}$ is experimentally demonstrated. It is experimentally verified that the period of the fringes generated in EPSLON configuration depends on $\lambda_{ex}$ and $\theta$ and the resolution of the final DL image depends in addition also on $\lambda_{det}$.

Next, the potential of EPSLON is demonstrated on 100 nm polystyrene beads, Fig. 3b. A straight waveguide is employed, and the images of beads placed directly on top of waveguide core are acquired using a detection MO with numerical aperture NA = 0.9. The acquired image stack of 100 frames is given as input to a fluorescence-based super-resolution algorithm called as Super-resolution method based on Auto-Correlation two-step Deconvolution (SACD) [9]. The choice of SACD over other IFON algorithms is based on the simulation studies shown in supplementary section 3 and due to better performance of the algorithm at low signal to background ratios [66]. Now to validate the super-resolved images generated by SACD, the same sample is imaged by a scanning electron microscope (SEM), which serves as the ground truth image. In Fig. 3b, the line profiles indicate the intensity variations across the particles in the insets. The line profile of the green inset clearly indicates that EPSLON resolves the unresolved nanobeads shown in the red inset in the DL image. The peak-to-peak distance between the beads is 180 nm. It is seen that EPSLON and the ground truth image (SEM) agree well.

Next, to showcase the potential of EPSLON for life sciences, we choose biological samples such as extra-cellular vesicles (EVs), human placental tissue and rat kidney sections. The first of them, e.g., small EVs, are gaining attention due to their role in intercellular communication and possible clinical applications, especially for targeted drug delivery. Nevertheless, their molecular biology, as well as their therapeutic potential, is far to be completely understood. Further understanding of the spatiotemporal aspects of EVs rely on the ability to image processes such as EV secretion, uptake and biodistribution. However, imaging and tracking of small EVs has been challenging due to their small sizes (50-200 nm), and often require the use of labeling strategies, that may alter EV release and structure, prior to visualization [67]. These problems are mitigated in EPSLON: the decoupled speckle-illumination and detection paths helps visualize these structures beyond the diffraction-limit with high-contrast and without photobleaching as demonstrated in Fig. 4a. The EVs used in this experiment have a size distribution of 75-250 nm, Supplementary Fig. 16, and are fluorescently labeled. EPSLON and total-internal reflection fluorescence (TIRF) imaging of the same region of interest is performed. Fluorescent dyes are chosen in a way to ensure that there is no fluorescent signal reaching the camera during EPSLON imaging. 50 images are acquired in both EPSLON and TIRF mode using a detection MO with NA = 0.45. It is seen in Fig. 4a and Fig. 4b that the EPSLON and TIRF images are in good agreement. The line profile in the red and green insets in Fig. 4b show a blob of light and therefore, unable to clearly resolve the EV particles. To resolve these particles, the DL image stack is given as input to the reconstruction algorithm to generate the EPSLON image. From Fig. 4b it is seen that in the EPSLON image the EVs are clearly resolved as shown by the line profiles. To validate the EPSLON result, the same region of interest is imaged with a higher NA = 0.9 detection MO. The EPSLON matches well with the line profile of the ground truth optical image.

Next, we demonstrate the utility of EPSLON for super-resolution imaging of kidney tissue. Particularly, the ultrastructural analysis of this organ is of great importance for the identification of several renal pathologies such as, for example, the minimal change disease [68]. While in recent years significant attention has been given to fluorescence-based super-resolution optical methods for kidney research and diagnosis [69], the exploration of label-free optical super-resolution approaches in these cases remain largely unexplored and this is why EPSLON is promising. In this work, we employed rat kidney sections as a test sample as shown in Fig. 5. 250 frames of rat kidney sections are imaged first in label-free mode, then in waveguide TIRF mode (i.e., using fluorescence markers) and finally using SEM. Fig. 5a shows the label-free DL and EPSLON images of a rat kidney section acquired using a detection MO with NA = 1.42. The central portion of the image labelled 'G' in yellow font represents a kidney glomerulus, while regions labelled 'PT' in yellow font correspond to the proximal tubuli in the kidney section. Three regions of interest labelled 'b', 'c' and 'd' in Fig. 5a are blown up and shown. Fig. 5b1 and Fig. 5c1 are TIRF-DL images of regions enclosed in 'b' and 'c'. Its corresponding TIRF-SACD images are shown in Fig. 5b2 and Fig. 5c2 respectively. Label-free DL images of regions enclosed in 'b' and 'c' are shown in Fig. 5b3 and Fig. 5c3. Its corresponding EPSLON images are shown in Fig. 5b4 and Fig. 5c4 respectively. Fig. 5d1 and Fig. 5d2 are TIRF-DL and TIRF-SACD images of the region labelled 'd'. Label-free DL and EPSLON images of the region labelled 'd' are shown in Fig. 5d3 and Fig. 5d4 respectively. A SEM, scanning electron microscope, image of the same region is shown in Fig. 5d5. It must be noted that EPSLON is an evanescent wave illumination technique and thereofore, EPSLON image shows structures which are within the penetration depth of the evanescent field, while SEM image showcases structures on the top of the sample, i.e., analogous to an epi-illumination scheme. Supplementary Movie 1 compares DL and EPSLON at different regions of interest in Fig. 5. The resolution of optical images acquired is quantified using FRC, Fourier Ring Correlation [71]. The mean resolution of label-free DL image is 256 nm and for EPSLON is 133 nm, which implies a ~1.92 fold resolution gain in the label-free mode. The local FRC plot of the whole field-of-view in Fig. 5 is shown in supplementary Fig. 17.

Another potential application of EPSLON in medical sciences is that it facilitates the screening of diseases that require high-resolution visualization over large areas for accurate diagnosis. An example of this is kidney histopathology, where light microscopy is first used for contextual examination of the tissue microanatomy, followed by electron microscopy to visualize ultrastructural details not visualizable via conventional optical microscopy. We demonstrate the applicability of EPSLON in clinical settings by analyzing a human kidney sample preserved by the standard formalin-fixation and paraffin-embedding methodology [72]. The mean FRC resolution of the EPSLON image is 189 nm. After identifying a glomerular area using bright field modality at low magnification, see inset in Fig. 6a, the region of interest was further inspected in EPSLON at 60X magnification using an M.O. with a NA = 0.9. The resulting image, shown in panel Fig. 6a, offers a large FOV high-contrast view of the intricate anatomy of the glomerulus and its surrounding tissue including the Bowman's capsule (BC), and many adjacent proximal tubuli (PT). A zoom-in visualization over a glomerular capillary, presented in Fig. 6b, allows for a clear observation of relevant structures such as the glomerular basement membrane (GBM) between a capillary lumen (CL) containing a red blood cell (RBC), and a surrounding podocyte cell with its distinctive nucleus (N). The GBM thickness is a key indicator for the diagnosis of several renal conditions such as thin basement membrane disease and Alport syndrome [73], involving GBM thinning, or primary membranous nephropathy [74], with GBM thickening. In clinical settings, this microanatomical feature is usually measured using ultrathin polymer-embedded tissue sections in transmission electron microscopy. In this example, a line profile measurement over the GBM reveals a lateral resolution down to 144 nm in EPSLON, blue line in Fig. 6b and 6d, which allows for an optical sub-diffraction limit measurement on a conventional paraffin-embedded sample. The GBM region is not discernible in the DL image counterpart, green line in Fig. 6c and 6d. Another region of interest in the same FOV is shown in supplementary Fig. 18. Finally in supplementary Fig. 19 and supplementary Fig. 20, we showcase the potential of EPSLON for the evaluation of human placental tissue sections using straight waveguides in tandem with SACD.

**Discussion and Outlook**

Fluorescence-based algorithms have been previously applied to coherently scattering specimens [47, 75]. However, the reconstructed images generated must be interpreted with caution with regards to resolution beyond the diffraction-limit as coherence of the scattered light cannot be neglected [13, 32]. In our proof-of-concept work EPSLON, we circumvent this issue for two-dimensional samples by realizing a light source whose electric field has δ-function correlations, i.e., by using the PL property of $Si_3N_4$ waveguides for near-field illumination of unlabeled samples. In addition, PL emission takes place within the core matrix and a part of it gets transmitted into the far-field, which prevents realizing an ideal chip-based imaging system where only the scattered light off the sample reaches the camera. Despite the limited photon budget, we have demonstrated proof-of-concept label-free super-resolution results on nanobeads, EVs, human placenta tissue and rat kidney sections. The experimental particulars are provided in Table 2 and Table 3 in the supplementary material. In future works, we aim for novel chip designs [76, 77] which will improve the signal to background ratio. The other challenge of lack of specificity in label-free imaging can be mitigated by resorting to machine learning based tools [78].

This work lays the foundation for synthesizing a label-free incoherent imaging system that is compatible with the myriad of fluorescence-based super-resolution algorithms to circumvent the spatial diffraction-limit. We believe that EPSLON will trigger further developments of label-free super-resolution incoherent optical microscopy methods and its application in biology, with particular attention to histological analyses where a fast, simple, cost-effective, and high-resolution imaging method is beneficial for medical guidance and diagnosis. Future studies will also investigate the role of intrinsic autofluorescence of tissue sections in label-free imaging. Expanding the concepts of EPSLON to be applied in tandem with other fluorescence-based super-resolution algorithms like Non-linear SIM [79] and STED [80] is currently in progress. This work could also initiate further developments within integrated optics to harness the PL properties of different materials. Interestingly, PL in waveguides is an undesirable phenomenon as it increases the propagation losses of the guiding structures. However, here we demonstrated that PL of $Si_3N_4$ waveguide can be harnessed for near-field illumination to develop an incoherent imaging system for surpassing the diffraction limit in label-free regime.

**Ethical statement**

Full-term placentae from 10 different Caucasian healthy patients were collected anonymously immediately after delivery at the University Hospital of North Norway. Written consent was obtained from the participants according to the protocol approved by the Regional Committee for Medical and Health Research Ethics of North Norway (REK Nord reference no. 2010/2058–4).

**Acknowledgements**

We wish to acknowledge the fruitful discussions with Prof. Olav Gaute Helleso (UiT), Dr. Florian Strohl (UiT) and Prof. Krishna Agarwal (UiT). Also acknowledgments to Prof. Ganesh Acharya and Associate Prof. Mona Nystad (UiT) for providing us with the tissue placenta sections and Assoc. Prof. Neoma Tove Boardman and Nirgun Basnet for providing rat kidney tissue sections. We are grateful to Fernando Cázarez-Márquez for assisting with the rat kidney embedding and sectioning.

This project has received funding from the European Union’s Horizon 2020 research and innovation program under the Marie Skłodowska-Curie Grant Agreement No. 766181, project “DeLIVER”. BSA acknowledges the funding from the Research Council of Norway, projects # NANO 2021–288565 and # BIOTEK 2021–285571, and from the European Innovation Council (EIC), EIC-Transition project # 101058016.

**Authors and Affiliations**

**Dept. of Physics and Technology, UiT The Arctic University of Norway, Tromsø 9037, Norway**

Nikhil Jayakumar, Luis E. Villegas-Hernández, Hong Mao Jean-Claude Tinguley, Balpreet Singh Ahluwalia

**Dept. of Microsystems and Nanotechnology, SINTEF Digital, Gaustadalleen 23C, 0373 Oslo, Norway**

Firehun T Dullo

**Dept. of Molecular Cell Biology, Institute for Cancer Research, Oslo University Hospital, Oslo, Norway**

Krizia Sagini, Alicia Llorente

**Centre for Cancer Cell Reprogramming, Faculty of Medicine, University of Oslo, Montebello, 0379 Oslo, Norway**

Krizia Sagini, Alicia Llorente

**Dept. for Mechanical, Electronics and Chemical Engineering, Oslo Metropolitan University, Oslo, Norway**

Alicia Llorente

**Innovation Photonics and Imaging Center, School of Instrumentation Science and Engineering, Harbin Institute of Technology, Harbin, China**
Weisong Zhao

## Author Contribution

NJ conceptualized the idea. NJ and BSA designed the experiments. WZ performed SACD and BlindSIM reconstructions. LEVH prepared the human placenta tissue and rat kidney sections on chip and imaged along with NJ. HM and LEVH performed SEM imaging of tissue sections. FTD and JCT designed the waveguide chip and mask for fabrication. FTD also performed the FIMMWAVE simulations. KS and AL provided the extra-cellular vesicles for imaging and wrote the EV preparation protocol. NJ performed the experiments, analyzed the results and prepared the manuscript with inputs from FTD, JCT, and BSA. NJ, LEVH, HM and JCT prepared the figures. LEVH prepared the video animations in the supplementary section. All authors commented on the manuscript. BSA secured the funding and supervised the project.

**Conflict of interest statement:** B.S.A. have applied for patent for chip-based optical nanoscopy. B.S.A is the co-founder of the company Chip NanoImaging AS, which commercializes on-chip super-resolution microscopy systems. All other authors declare no conflicts of interest regarding this article.

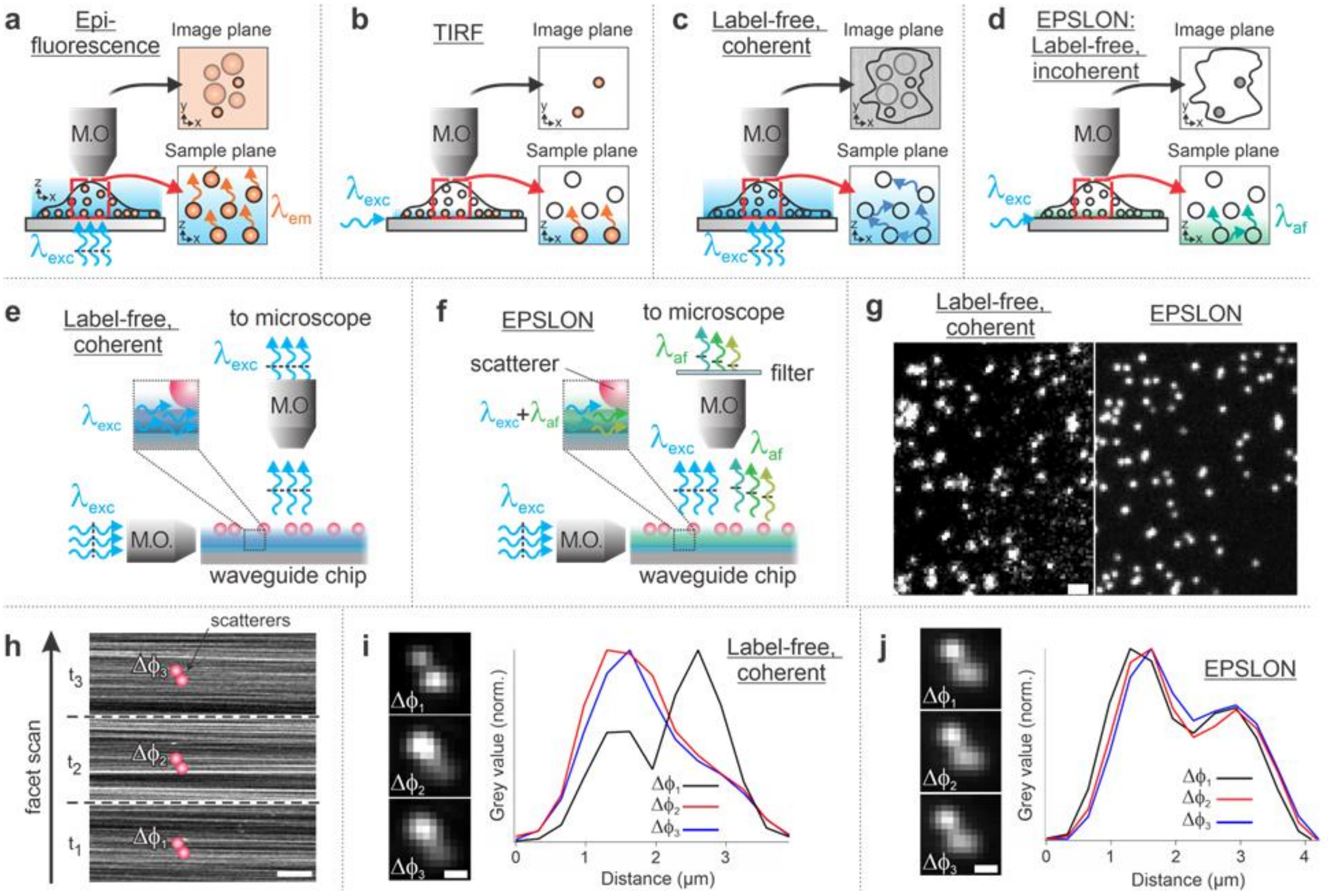

**Figure 1**: **Overview of ELSPON.** (**a-d**) Comparison between fluorescence and label-free microscopy. (**a**) Epifluorescence: coherent light $\lambda_{exc}$ is used for excitation of fluorescent molecules and the camera detects the Stoke shifted incoherent light emitted by the molecules $\lambda_{em}$. Stochastic fluctuations of the Stoke shifted light and specificity offered by the molecules helps suppress speckle noise enabling high-contrast imaging. (**b**) TIRF: coherent light for near-field illumination of fluorescently labeled samples and incoherent light gets detected by the camera. Near-field illumination helps to further improve the contrast as compared to the epifluorescence by illuminating thin sections of the sample. (**c**) Label-free coherent imaging: coherent light for illumination and coherent light gets detected by the camera. Multiple scattering and coherent nature of the scattered light leads to speckle noise, (**d**) EPSLON: incoherent light for near-field illumination of unlabeled samples and incoherent light scattered by the sample, $\lambda_{af}$, forms the image. The incoherent nature of the detected light in addition to the near-field illumination helps generate high-contrast label-free images. (**e**) Schematic of optical waveguide-based label-free coherent imaging. The guided coherent light generates an evanescent field that interacts with the sample placed at the core-cladding interface,. (**f**) Schematic of EPSLON imaging using $Si_3N_4$ waveguide. The guided coherent light induces incoherent photoluminescence (PL) in the core of the waveguide that interacts with the sample and gets transmitted into the far-field. (**g**) 200 nm gold nanoparticles imaged in coherent and EPSLON mode, scale bar 100 µm. The issues of coherent noise, poor-contrast associated with conventional label-free techniques is mitigated in EPSLON due to δ-function correlations existing in the detected light. (**h**) To induce fluctuations in image intensity over time, different modes of the waveguide are excited by scanning the coupling objective along the input facet of the waveguide. At each instance of time $t_1$, $t_2$, $t_3$ etc. the scatterers get excited by different MMI patterns, and an image is acquired. Scale bar 10 µm. (**i**) Experimental demonstration of coherence of scattered light using two 200 nm gold nanoparticles, scale bar 10 µm. The excitation of different modes causes the phase difference $\Delta\varphi$ between the scattered light off the particles to change, leading to different images at different instances of time in label-free coherent imaging. This is demonstrated by the line profile of the bead images provided alongside. (**j**) Experimental realization of stochastic nature of the detected light in label-free imaging. The same nanoparticles shown in (i) are imaged in EPSLON mode, scale bar 10 µm. In EPSLON, stochastic fluctuations between the scattered incoherent PL light reaching the camera leads to identical images at different instances of time. This can be seen from the line profiles of the bead images provided alongside.

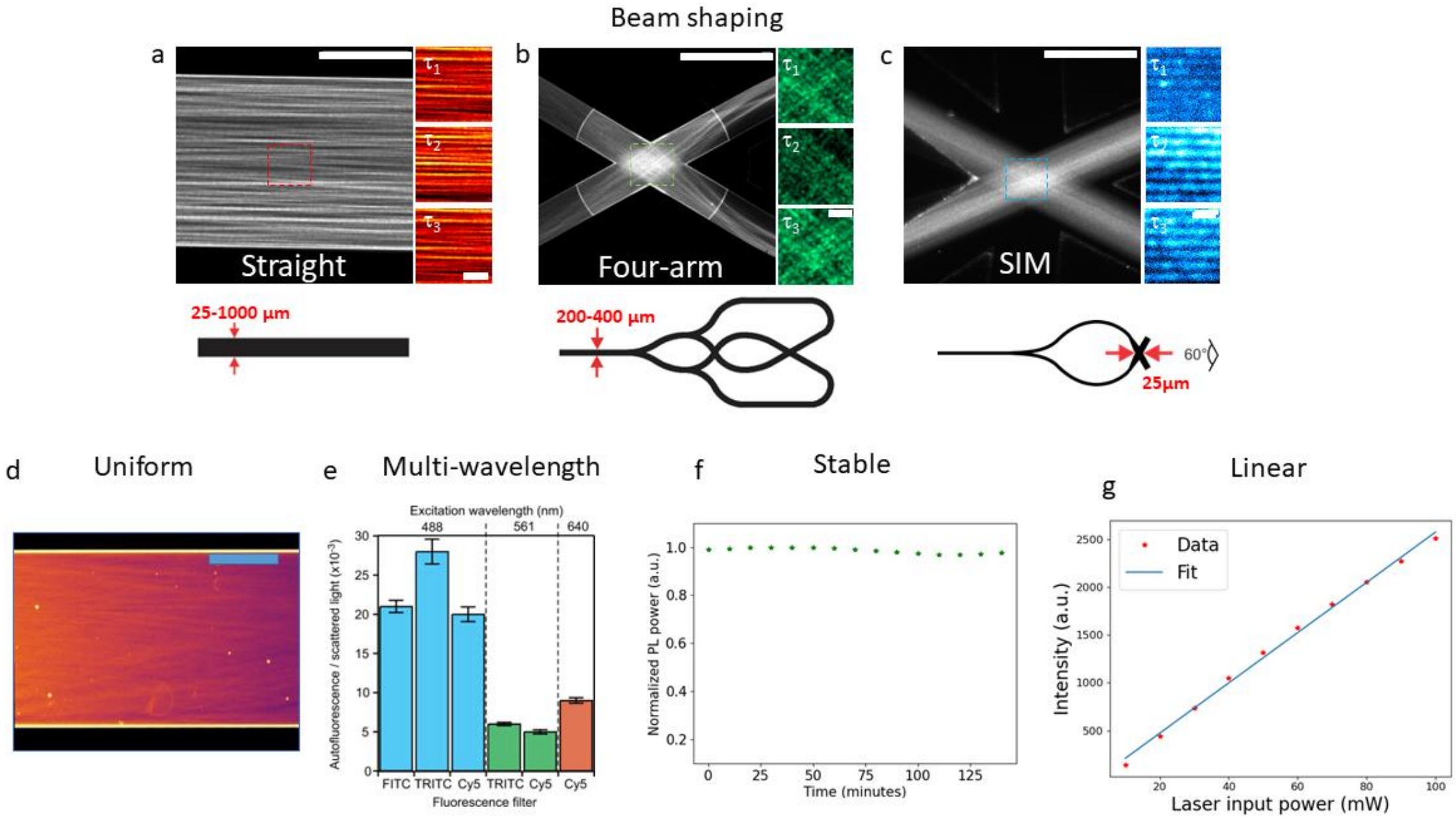

**Figure 2: Structuring the incoherent photoluminescent illumination using $Si_3N_4$ waveguides and its properties.** (**a-c**) Structuring photoluminescence using different waveguide geometries. (**a**) Straight waveguide that supports multiple modes in the core, scale bar 20 µm. The red insets are a part of the imaging region on the straight waveguide that is blown up to show the MMI pattern at different instances of time, scale bar 5 µm. Schematic diagram shows the geometry and width of the straight waveguide fabricated on a wafer. (**b**) Four-arm crossing waveguide provides more illumination spatial frequencies, scale bar 20 µm. The green inset is the blown-up region of the imaging area on the four-arm crossing waveguide, showing different speckle patterns at different instances of time, scale bar 5 µm. Schematic diagram shows the geometry and width of the four-arm crossing waveguide. (**c**) SIM chip where two single-mode waveguides are made to overlap at the imaging area enclosed by the blue inset, scale bar 20 µm. The blown-up regions show the interference fringe pattern at three different phases, 2 µm. Schematic diagram shows the geometry of the SIM chip for two-dimensional SIM works. (**d**) By averaging out several MMI patterns, an illumination profile devoid of speckle patterns, as shown in supplementary Fig. 2b, over large field-of-view can be generated, scale bar 100 µm. The color grading visible from left to right is attributed to propagation loss (Supplementary Table 1). (**e**) Ratio of incoherent PL scattering to coherent scattering for a single waveguide at three different wavelengths is plotted. Here the waveguide is excited at 488 nm, 561 nm and 640 nm and the corresponding emissions may be detected in FITC, TRITC and CY5 channels. (**f**) Normalized PL emission with a root mean squared value of 1007 a.u. is plotted here as a function of time. No bleaching was observed for about 2 hours of imaging indicating a stable emission that is suitable for long-term cell imaging. (**g**) PL emission varies linearly with the input coupling power, implying a linear system.

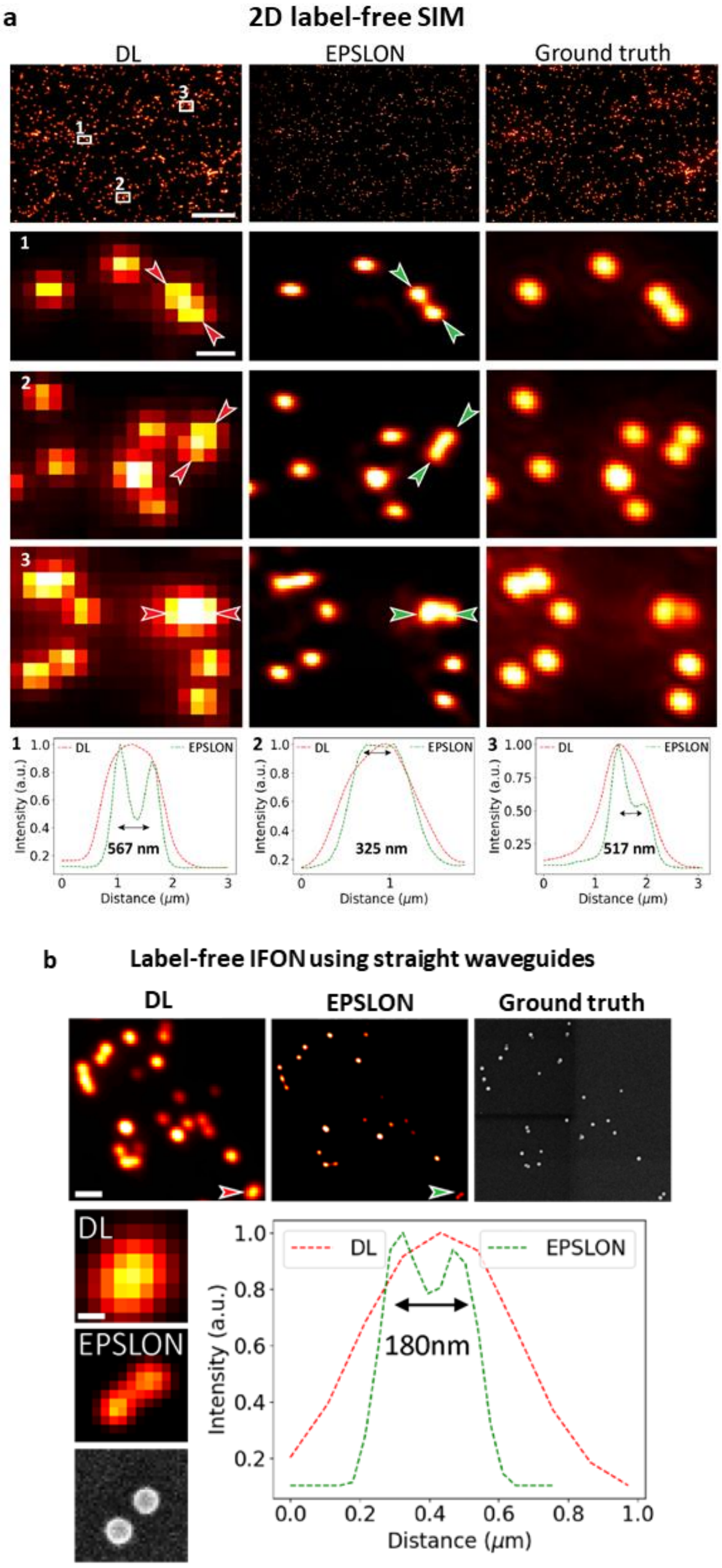

**Figure 3: (a) EPSLON via BlindSIM for label-free 2D SIM of 195 nm polystyrene beads using four-arm junction waveguide**. Large field-of-view diffraction-limited (DL), super-resolved (EPSLON) and ground-truth images are shown, scale bar 20 µm. The DL images are acquired using 0.75 NA and ground truth optical image with 60X/1.2 NA objective lens. Three regions of interest enclosed in white boxes in DL image are blown up and shown, scale bar 1 µm. The corresponding EPSLON and ground-truth large images are also blown up and shown alongside. The red arrowheads indicate unresolved beads in the DL images and the green arrowheads indicate the resolved beads in the EPSLON images. The line plots corresponding to normalized intensity variations along these arrows show the resolution improvement in EPSLON. **(b) EPSLON via SACD for label-free super-resolution imaging of 100 nm polystyrene beads using straight waveguides**. Diffraction-limited DL image, super-resolved EPSLON image and scanning electron microscope (SEM) ground truth image are shown, scale bar 1 µm. The red and green line plot corresponds to intensity variation along the red and green arrows in the DL and EPSLON images respectively, scale bar 100 nm.

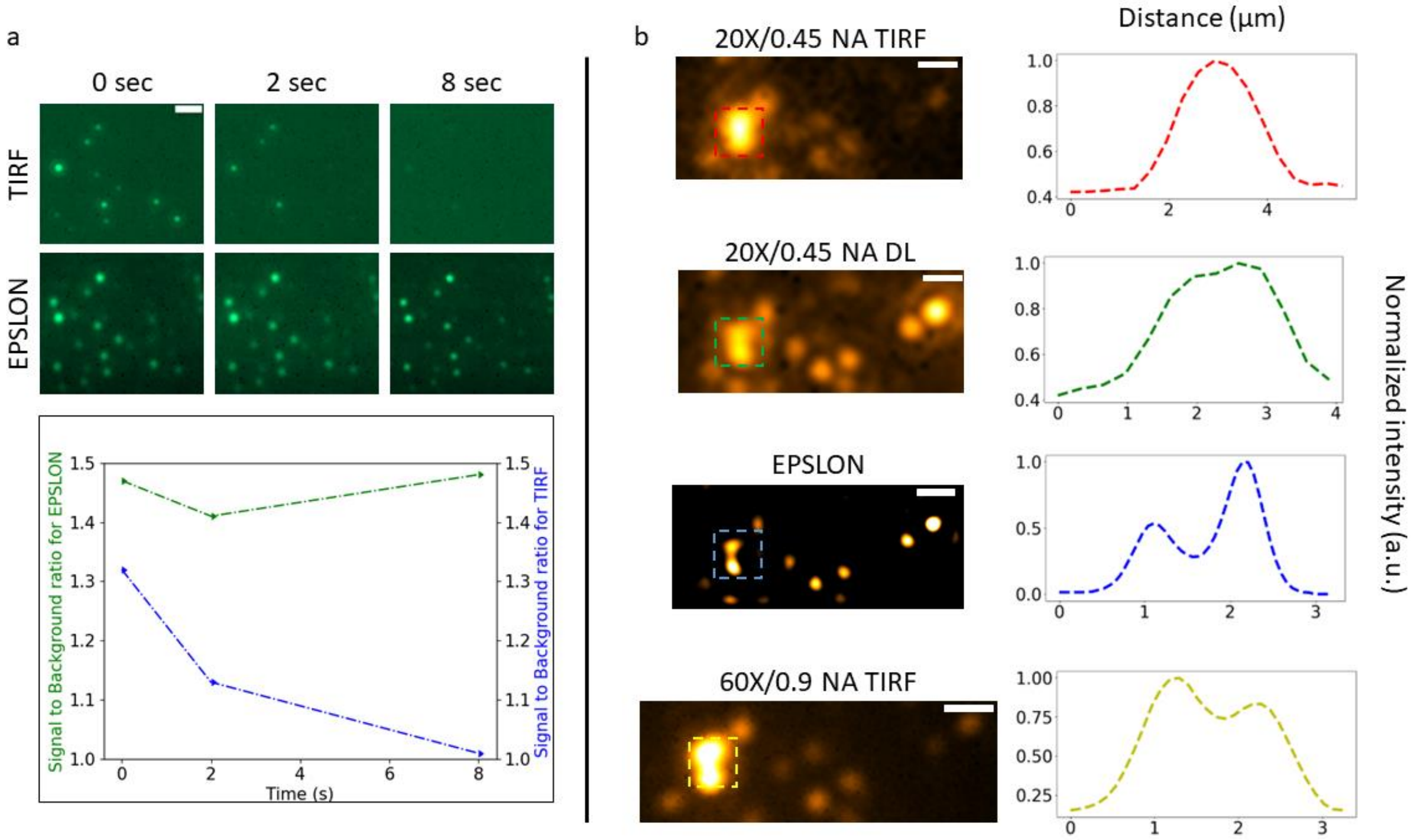

**Figure 4: EPSLON via SACD for circumventing photobleaching and for label-free super-resolution imaging of EVs.** (**a**) Time-lapse imaging comparison of extra-cellular vesicles between TIRF and EPSLON configurations. EPSLON helps to image nanosized EVs over long periods of time without photobleaching and with better signal-to-background ratio as opposed to TIRF, scale bar 5 µm. This fact is quantified in the graph where signal-to-background ratio as a function of time is plotted. (**b**) Super-resolution imaging of EVs in label-free regime using EPSLON configuration. EVs are imaged in both TIRF and EPSLON mode, scale bar 2 µm. The red, green, blue, and yellow insets correspond to EVs in diffraction-limited TIRF image, label-free diffraction-limited image termed DL, label-free super-resolved EPSLON image and TIRF ground truth image. The line profiles corresponding to each of these insets showing the intensity variation are also shown alongside. EPSLON resolves the unresolved EVs in the diffraction-limited images and this result is validated by the TIRF ground truth image acquired with a higher NA objective.

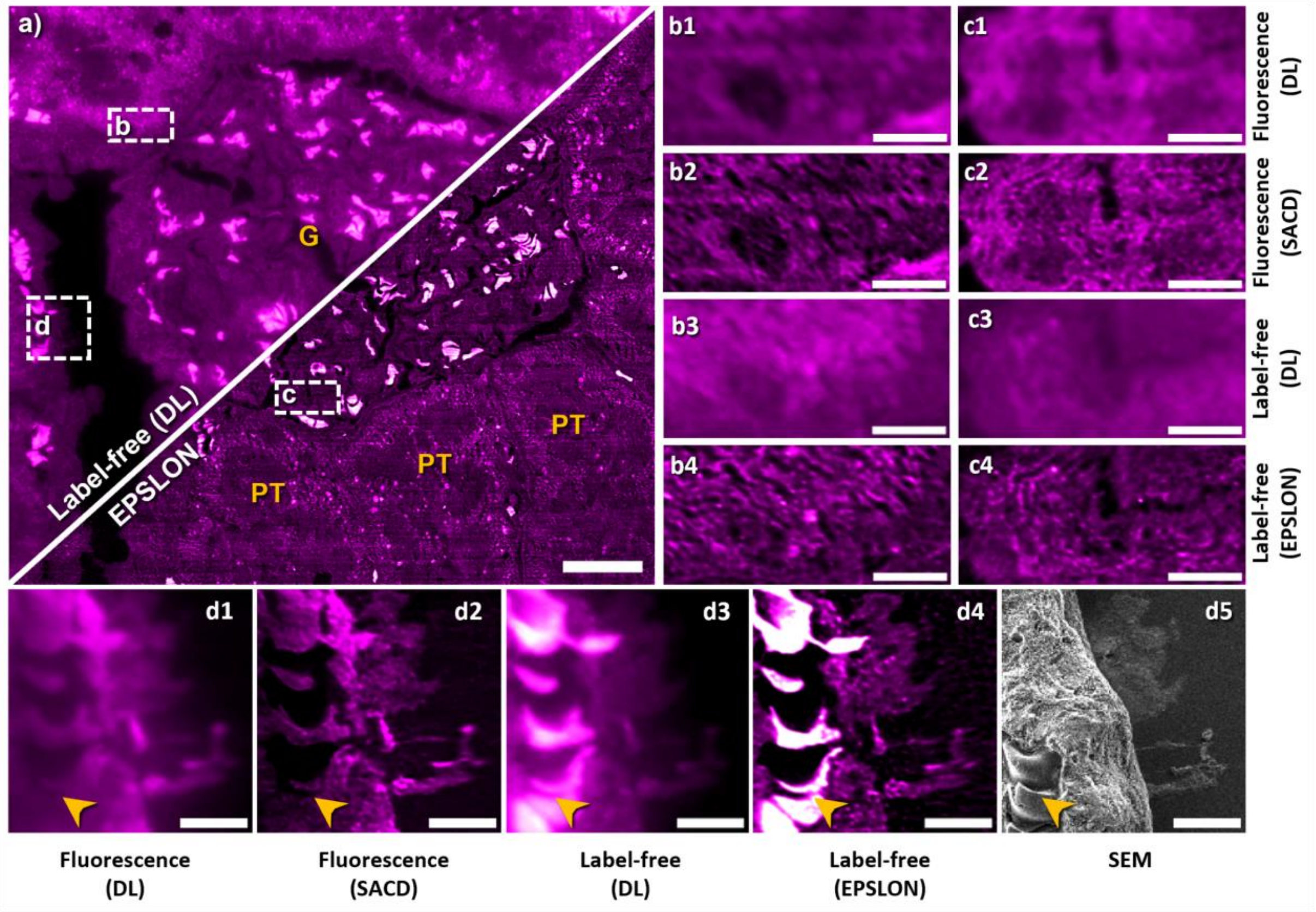

**Figure 5: EPSLON via SACD for label-free super-resolution imaging of rat kidney tissue sections and benchmarking via correlative microscopy, EPSLON-TIRF and ELSON-SEM. (a)** Label-free diffraction-limited (DL) and its corresponding super-resolved EPSLON images are shown. The region 'G' labelled in yellow font indicates Glomerulus and the regions labelled 'PT' in yellow font indicate Proximal Tubuli of the kidney section. Three regions of interest enclosed in white dotted boxes are labelled 'b', 'c' and 'd'. Scale bar 25 µm. **(b1, c1, d1)** Diffraction-limited TIRF images of the regions enclosed by white dotted boxes in (a) are blown up and shown, scale bar 5 µm. **(b2, c2, d2,)** Corresponding super-resolved TIRF images of regions in (b1, c1, d1) are shown, scale bar 5 µm. **(b3, c3, d3)** Label-free DL images of regions enclosed by the white dotted boxes in (a) are shown magnified, scale bar 5 µm. **(b4, c4, d4)** Super-resolved EPSLON images of regions corresponding to (b3, c3, d3) are shown, scale bar 5 µm. **(d5)** SEM image of the region enclosed by the white dotted box labelled 'd' is shown. The yellow arrowheads denote the location of red blood cells, being imaged throughout all the microscopy methods. Scale bar 5 µm. The mean FRC resolution of label-free DL image is 256 nm and the mean FRC resolution of EPSLON image is 133 nm.

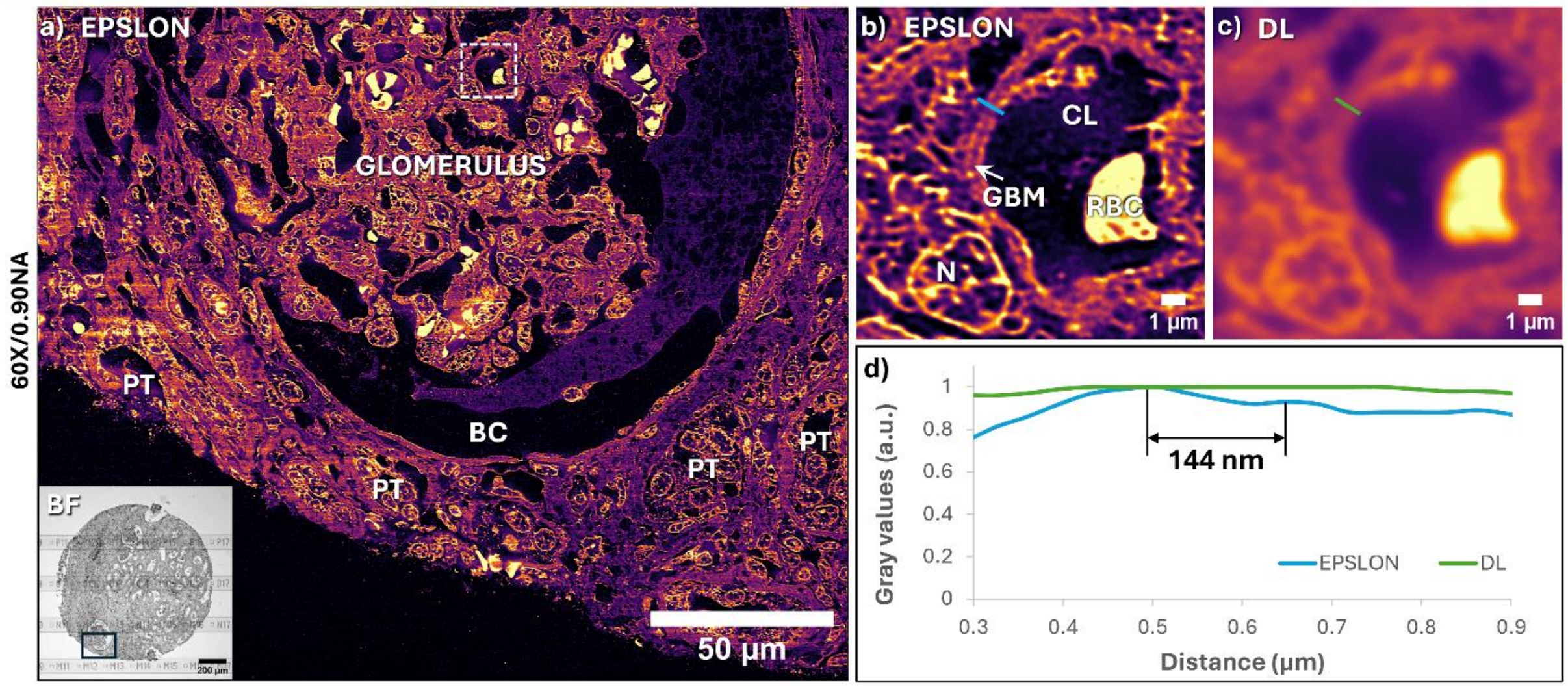

**Figure 6: EPSLON via SACD for human kidney histopathology. (a)** Lable-free super-resolved EPSLON image of a human kidney tissue section is shown, scale bar 50 µm. A mean FRC resolution of 129 nm is observed. The gray scale inset is the bright field image of the kidney section imaged with a 4X/0.10 NA detection objective, scale bar 200 µm. The pseudo-colour image illustrates the EPSLON results of the selected glomerular region in the BF image enclosed within the black box. EPSLON provides a high-contrast contextual visualization of the sample, enabling the identification of structures such as the glomerulus, the Bowman's capsule (BC), and many adjacent proximal tubuli (PT). **(b)** The region in the EPSLON image enclosed within the white dotted box is blown-up and shown, scale bar 1µm. This figure reveals microanatomical structures including a glomerular basement membrane (GBM), a capillary lumen (CL) with a red blood cell (RBC), and the nucleus (N) of an adjacent podocyte. **(c)** The corresponding diffraction-limited (DL) view of the FOV in (b) is shown, scale bar 1µm. Note the poor contrast and resolution as compared to the EPSLON image. **(d)** Line profile measurements over the GBM reveal a lateral resolution of approximately 144 nm using EPSLON, which cannot be measured in DL modality. The blue line represents intensity variation in the EPSLON image along the region shown in (b) and green line represents intensity variation along the DL region shown in (c).

# Label-free incoherent super-resolution optical microscopy

Nikhil Jayakumar**[1], Luis E. Villegas-Hernández[1], Weisong Zhao[2], Hong Mao[1], Firehun T Dullo[3], Jean-Claude Tinguley[1], Krizia Sagini[4,5], Alicia Llorente[4,5,6], Balpreet Singh Ahluwalia*[1,7]

[1]*Department of Physics and Technology, UiT The Arctic University of Norway, Tromsø, 9037, Norway*
[2]*Innovation Photonics and Imaging Center, School of Instrumentation Science and Engineering, Harbin Institute of Technology, Harbin, China*
[3]*Department of Microsystems and Nanotechnology, SINTEF Digital, Gaustadalleen 23C, 0373 Oslo, Norway*
[4] *Department of Molecular Cell Biology, Institute for Cancer Research, Oslo University Hospital, The Norwegian Radium Hospital, 0379 Oslo, Norway*
[5]*Centre for Cancer Cell Reprogramming, Faculty of Medicine, University of Oslo, Montebello, 0379 Oslo, Norway*
[6]*Department for Mechanical, Electronics and Chemical Engineering, Oslo Metropolitan University, Oslo, Norway*
[7]*Department of Clinical Science, Intervention and Technology, Karolinska Institute, Sweden*

*Balpreet.singh.ahluwalia@uit.no, **nik.jay.hil@gmail.com

## 1. Label-free high-resolution microscopic techniques

Below we have categorized the state-of-the-art high-resolution label-free techniques into three categories:
Technique 1: Techniques that use the concept of synthetic aperture/spatial frequency shift for coherently scattering samples
Technique 2: Synthetic aperture/Spatial-frequency shift concepts for coherently scattering samples using chip-based solutions
Technique 3: Techniques that apply fluorescence-based algorithms to coherently scattering samples.

Technique 1: High-resolution techniques that use the concept of synthetic aperture for label-free microscopy

Abbe's resolution-limit when considering oblique illumination for elastically scattered light is $\frac{\lambda_{ill}}{NA_{ill}+NA_{det}}$. And in fluorescent microscopy, because of the emission properties of fluorescent molecules, the Abbe resolution limit is $\frac{\lambda_{det}}{2NA_{det}}$. Below we highlight the state-of-the-art label-free optical techniques and briefly elucidate their working principles. This will help to understand the differences between EPSLON and label-free state-of-the-art optical techniques.

Ref. [a]: different holograms corresponding to different illumination directions on the sample plane are recorded. After post-processing, a high-resolution image is obtained which is diffraction-limited in Abbe's sense. The best theoretical resolution after post-processing will be $\lambda_{ill}/(NA_{illu} + NA_{det})$, where $\lambda$ is the wavelength of the detected light, $NA_{illu}$ is the numerical aperture of the illumination objective and $NA_{det}$ is the numerical aperture of the detection objective.

Ref. [b]: typically, in Fourier Ptychography, an LED array is used to provide oblique illumination at the sample plane. Then using a phase retrieval algorithm from the intensity images, a high-resolution final image is generated. The final image is still limited by Abbe's diffraction limit.

Ref. [c]: several azimuthally varying illuminations at the sample plane is provided to perform synthetic aperture reconstruction of the sample. The final resolution is determined by Abbe's diffraction limit.

Ref. [d]: a $2\pi$ azimuthal scan of a laser beam at the back-focal plane of the illumination objective within the integration time of the camera generates a high-contrast image. The final resolution is limited by Abbe's diffraction limit.

Ref. [e]: sample is illuminated via a series of oblique illuminations. The scattered light off the sample and specular reflection from the $SiO_2$ substrate helps create a common-path interferometry configuration. Via post-processing a two times resolution over head-on illumination $\lambda_{ill}/NA_{det}$ is obtained. The final resolution is within Abbe's diffraction limit.

Techniques 2: High-resolution label-free chip-based techniques with concepts like synthetic aperture microscopy
The concepts elaborated in Technique 1 are translated to chip-based platforms here. Hence, an even higher resolution in principle is possible due to the higher refractive-index of the material employed. The resolution achievable is given by Abbe, $\frac{\lambda_{ill}}{NA_{ill}+NA_{det}}$. All these techniques mitigate coherent speckle noise, either by summing up different speckle patterns on intensity-basis at the camera plane or by using a broadband light source

Ref. [a]: samples are illuminated in multiple azimuthal directions via evanescent waves generated by a $Si_3N_4$ waveguide. A post-processing algorithm is then applied to generate a high-resolution label-free image. This resolution is determined by Abbe's diffraction limit of $\lambda_{ill}/(NA_{illu} + NA_{det})$.

Ref. [b]: In this work, broadband light emitted by the nanowire ring (NWR) is coupled into the film waveguide via a single mode. The shortest Stoke shifted wavelength emitted by the NWR will then determine the smallest coherence length. What it essentially implies is that the scattered light will have a constant phase relationship between different locations excited by this mode. Or in other words, the phase information is still preserved in the scattered light. Hence, in principle this technique is not suitable to be applied to in tandem with fluorescence based super-resolution algorithms like structured-illumination microscopy. This is contrary to EPSLON where incoherent imaging system is proposed.

Ref. [c]: a photonic-chip made of Gallium Phosphide (refractive index > 3) is used in this work. Evanescent waves are used for illuminating the sample. Using a reconstruction algorithm a very high-resolution image limited by Abbe's diffraction-limit can be generated. However, the technique is still coherent in nature and therefore, fluorescence based super-resolution algorithms cannot be applied to circumvent the far-field diffraction-limit. This paper demonstrates $\boldsymbol{\lambda}/4.7$ resolution in the label-free mode. Here, authors used objective lens of $N.A._{det}= 1.49$ and GaP chip is used for illumination, i.e., refractive index 3.3 = $N.A._{illum}$. Then Abbe's resolution limit is $\lambda_{ill}/(NA_{illu} + NA_{det}) = \boldsymbol{\lambda}_{ill}/4.79$.

Ref. [d]: This is also a coherent imaging technique where the sample is illuminated from multiple directions using evanescent waves generated by a waveguide. The broadband light emitted by F8BT is efficiently coupled into waveguides. The guided light then provides evanescent wave illumination for the sample. By illuminating the sample from multiple orientations, a high-resolution image is generated. The resolution of the final image is still limited by Abbe's diffraction limit. This paper demonstrates $\boldsymbol{\lambda}/3$ resolution in the label-free mode. Here, authors used objective lens of $N.A._{det}= 0.85$ and $TiO_2$ chip is used for illumination, i.e. refractive index 2.2 = $N.A._{illum}$. Abbe's resolution limit is therefore $\lambda/(NA_{illu} + NA_{det}) = \boldsymbol{\lambda}/3.05$.

Techniques 3: Techniques that apply fluorescence-based algorithms to coherently scattering samples.

a. Lee, Yeon Ui, et al. "Hyperbolic material enhanced scattering nanoscopy for label-free super-resolution imaging." *Nature communications* 13.1 (2022): 6631.

Here, fluorescence-based super-resolution algorithm BlindSIM is applied in label-free mode. In this work, organic hyperbolic materials are employed to create sub-diffraction sized speckles. The illuminating field is passed through a vibrating multi-mode fiber to create temporal variations. Multiple oblique illuminations in the azimuthal plane are applied at the sample plane. Then Blind-SIM is applied to generate a super-resolved image. The technique resolution is given by Abbe's limit for elastically scattered light, $\lambda_{ill}/(NA_{ill} + NA_{det})$. Here $NA_{ill}$ is exceptionally high due to the high-index organic hyperbolic material used which leads to a super-resolution. A true incoherent system is imperative to avoid the caveats linked to the application of fluorescence based super-resolution algorithms for the coherent imaging [Wicker, Kai, and Rainer Heintzmann. *Nature Photonics* 8.5 (2014): 342-344, Jayakumar et al. *Nanophotonics* 11.15 (2022): 3421-3436].

b. Dong, Biqin, et al. "Superresolution intrinsic fluorescence imaging of chromatin utilizing native, unmodified nucleic acids for contrast." *Proceedings of the National Academy of Sciences* 113.35 (2016): 9716-9721.

In this work, the intrinsic autofluorescence of the cells mimic the stochastic fluctuations from a fluorescent molecule and therefore, a fluorescence-based localization technique is applied to circumvent the diffraction-limit. This work depends on samples that have unique intrinsic autofluorescence and it is thus not universal to diversified specimens.

The novelty in EPSLON is in using incoherent point-like light sources for near-field illumination of unlabeled samples, i.e., photoluminescence of $Si_3N_4$. It implies that each location of the sample, within the penetration depth of the incoherent evanescent field, scatters the non-propagating incoherent field into the far-field. This permits the application of fluorescence-based algorithms like SIM and IFON in the label-free regime to circumvent Abbe's diffraction-limit. Such an illumination technique is imperative to avoid the caveats linked with coherent imaging [Wicker, Kai, and Rainer Heintzmann. *Nature Photonics* 8.5 (2014): 342-344, Jayakumar et al. *Nanophotonics* 11.15 (2022): 3421-3436].

## 2. Waveguide fabrication

A 2 µm thick oxide layer was thermally grown on a silicon wafer, followed by the deposition of 150 nm thick Si3N4 layer using plasma enhanced chemical vapor deposition (PECVD). The 2D channel waveguides were defined by photolithography process and etched using reactive ion etching (RIE). Then, silicon oxide layer was deposited using PECVD on the patterned nitride layer for protection. Finally, the oxide layer was patterned and removed from certain regions of each waveguide to create the imaging regions. The oxide layer was removed using combination of both dry and wet etching [1]. Thus, at the imaging region the specimen is in direct contact with the waveguide core layer, accessing the evanescent field.

### Waveguide modes

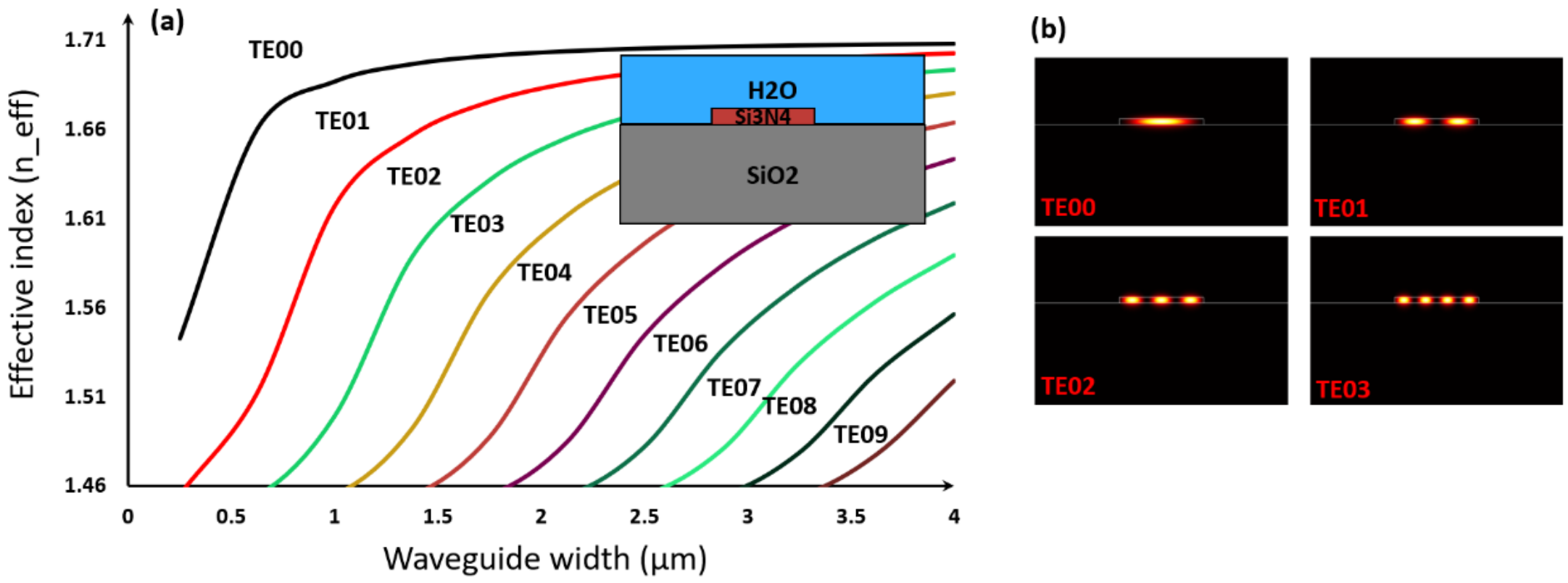

**Figure. 1**: (a) Effective indices of the guided modes for various $Si_3N_4$ waveguide widths. A schematic diagram of the cross-section of the waveguide is provided as an inset in the plot. (b) Mode profiles for the fundamental and higher-order TE - modes. The geometry of the waveguide in the simulation model is 150 nm thick and 0.25 - 4 µm wide. The guided modes for a strip Si3N4 waveguide were simulated using the commercial software FIMMWAVE (Photon Design, Oxford, UK), and its effective indices were calculated using the full-vectorial film mode matching (FMM) method.

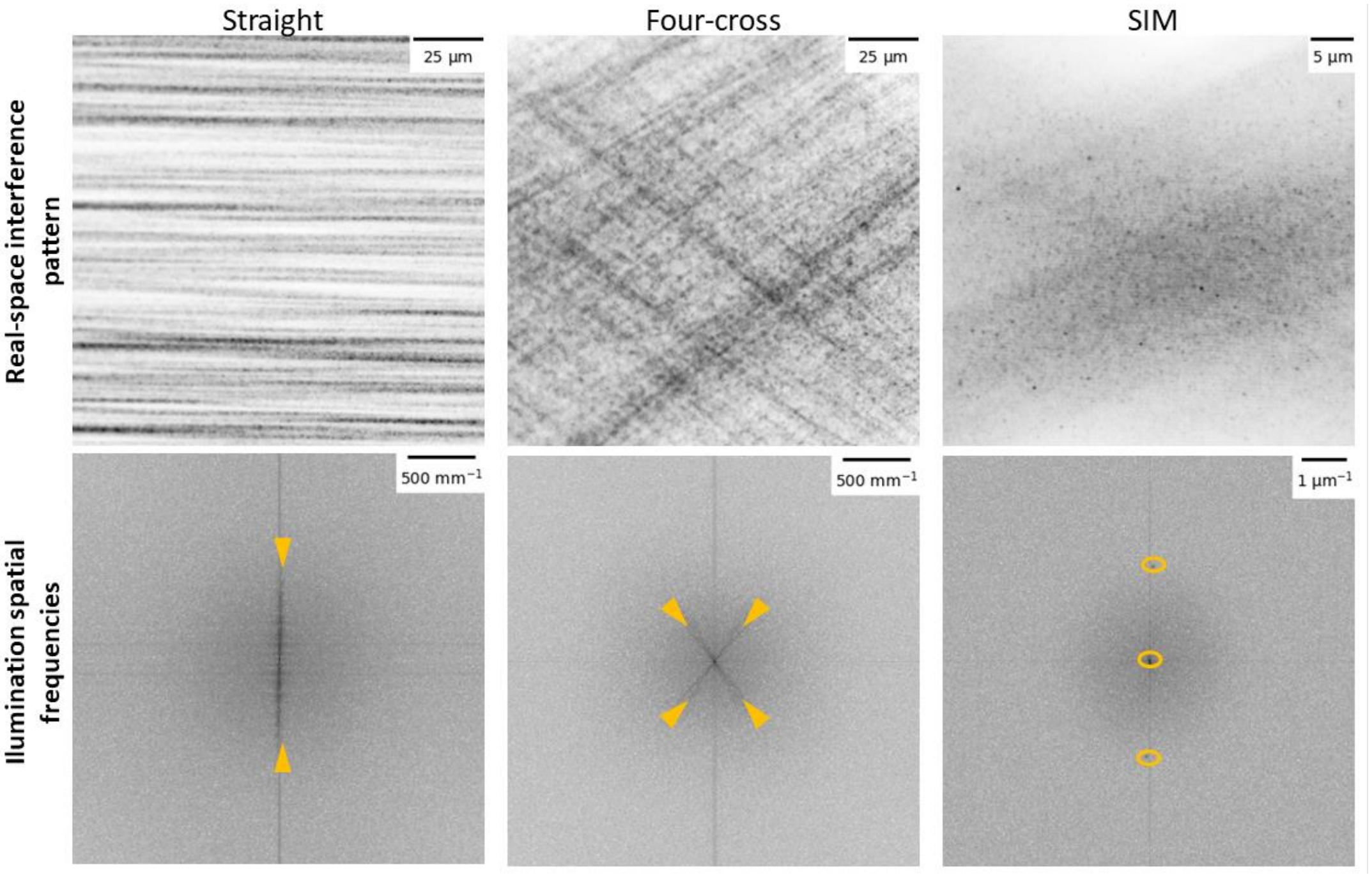

**Figure. 2a: Waveguide geometries employed in EPSLON for beam shaping and their corresponding illumination frequencies.** The multi-moded interference (MMI) pattern of straight and four-crossing waveguide is captured using a 20X/0.45 NA objective, scale bar 25 µm, while that of the SIM chip is captured using a 60X/1.2 NA objective, scale bar 5 µm. To see the well-defined fringe patterns in case of SIM chips, the interference angle between the overlapping single moded waveguides is chosen to be 20 degrees. The orange markers on the figures indicate the extend and orientation of the illuminating frequencies. Scale bar 500 mm$^{-1}$ for the straight and four-arm junction waveguide and 1 µm$^{-1}$ for the SIM chip.

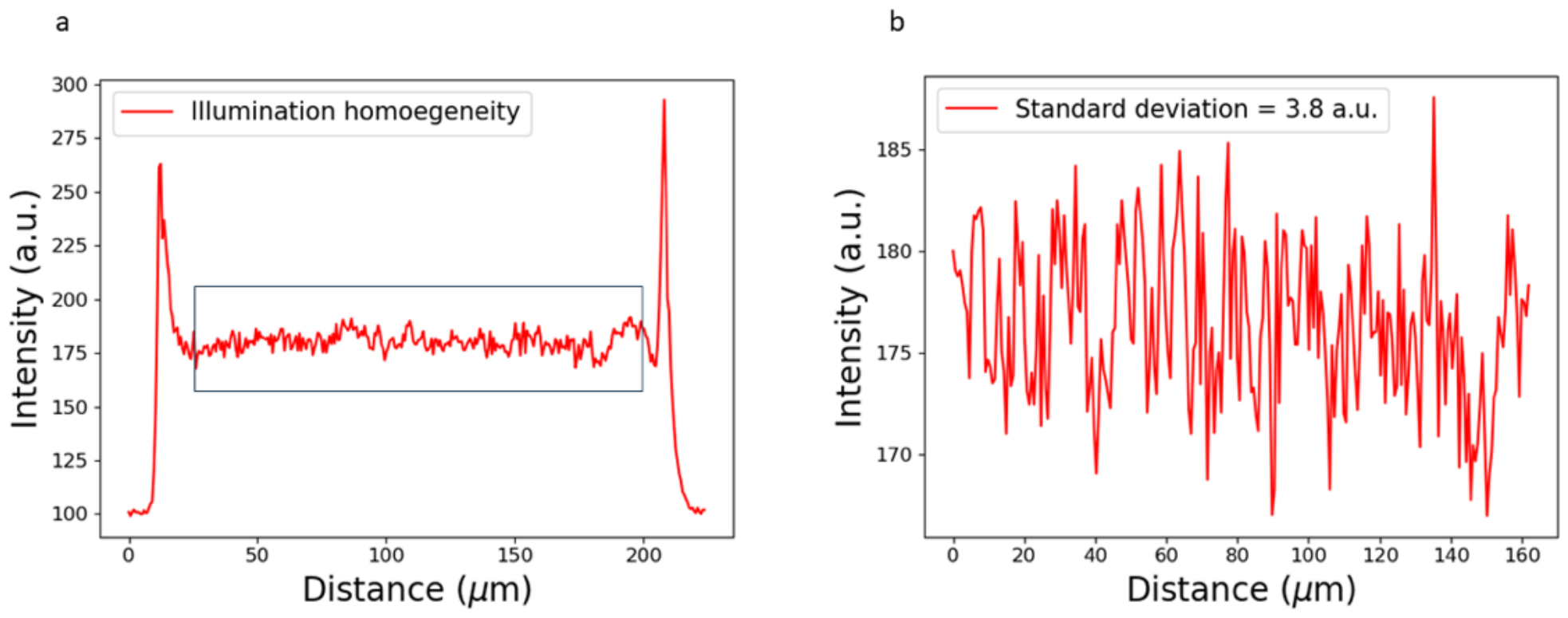

**Figure. 2b:** (a) Intensity variation along the width of a 200 µm waveguide is plotted. The two peaks in the plot correspond to the edges of the waveguide. (b) The black box in (a) is zoomed in and shown for better visualization purposes. The standard deviation of intensity over approximately 160 µm is 3.8 a.u.

**Table 1:**

| **Sl. No.** | **Excitation wavelength** | **Propagation loss** |
|---|---|---|
| 1 | 488 nm | ≈ 10 dB/cm |
| 2 | 561 nm | ≈ 2.5 dB/cm |
| 3 | 660 nm | ≈ 1dB/cm |

Table1: Propagation loss of a straight waveguide at different excitation wavelengths.

## 3. Simulation study and experimental verification of influence of multi-mode illumination patterns for usage in IFON

The following simulation studies elucidate how the multi-moded illumination pattern-induced correlation is mitigated to generate the super-resolved images in EPSLON.

Simulation1: The raw-image stack consists of three synthetic ring-like structures (220 nm, 260 nm and 300 nm from left to right). The data stack is then convolved with a PSF (220 nm) and down-sampled six times. Various types of noise (mixture noise with cytosol background, Poisson noise, Gaussian noise, out-of-focus light and baseline background) are then added to generate the final raw image stack. The emitters have an on/off fluctuating behavior. The fluctuation rate (On Time $\zeta_{on}$/Off Time $\zeta_{off}$) is to $\zeta_{on} = 1.67 \times 200$ frames and $\zeta_{off} = 2.5 \times 200$ frames, i.e., $\zeta_{on}/\zeta_{off} \approx 2/3$. The label density is set as 10000/μm$^2$ and all emitters have the same intrinsic brightness. An image stack of 100 frames is then reconstructed using 2$^{nd}$ order SOFI (SOFI$^2$) and 2$^{nd}$ order SACD (SACD$^2$). The structural similarity (SSIM) score is provided alongside each of the reconstructions as well.

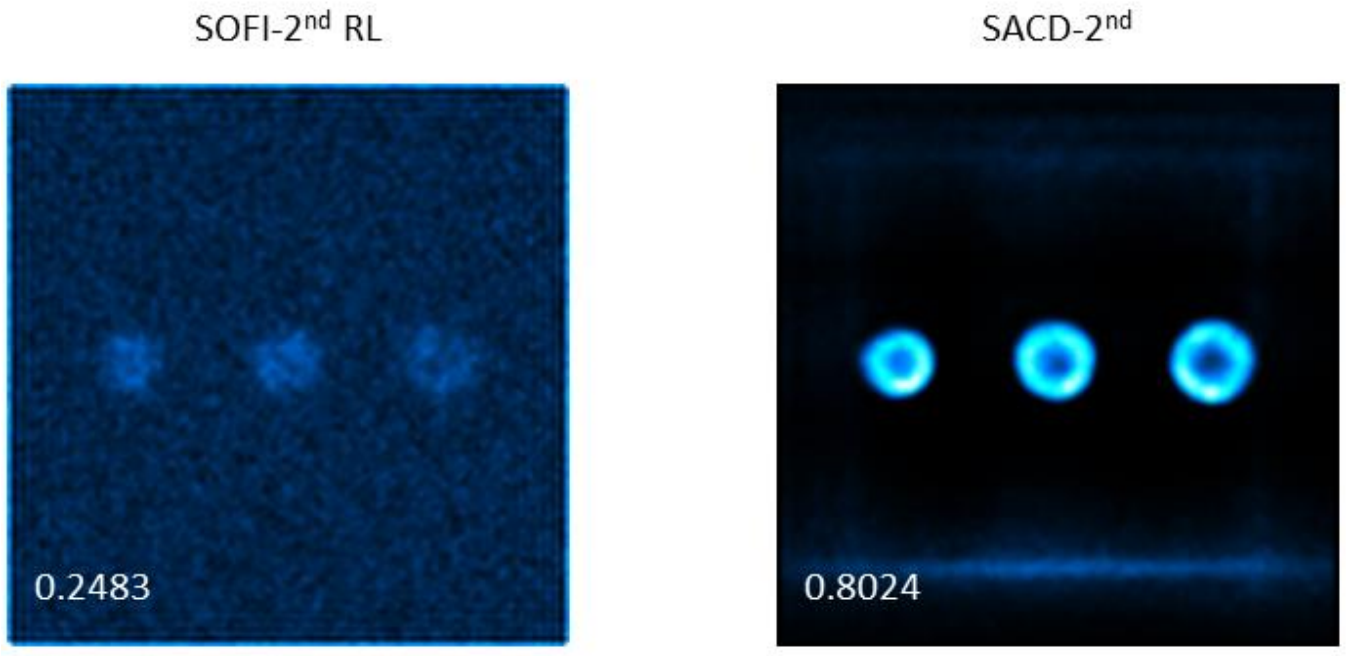

**Figure. 3**: Comparison of reconstructions using 2$^{nd}$ order SOFI and 2$^{nd}$ order SACD on the synthetic data set having slow intrinsic fluctuations ($\zeta_{on} = 1.67 \times 200$ frames and $\zeta_{off} = 2.5 \times 200$ frames).

It is seen that SACD$^2$ has been able to achieve a higher SSIM score than SOFI$^2$ for the slow fluctuation rate scenario we have chosen here, Fig. 3. The higher SSIM score of SACD is attributed to the pre-deconvolution steps involved in its reconstruction pipeline. This behavior is documented in Ref. [Zhao, Weisong, et al. "Enhanced detection of fluorescence fluctuations for high-throughput super-resolution imaging." *Nature Photonics* (2023): 1-8.] where SACD$^2$ is shown to have a high convergence even with just 20 frames.

Simulation2: The difference as opposed to Simulation1 is that the raw image stack is now multiplied with multi-moded illumination pattern. This will correspond to the situation where these very slowly fluctuating emitters are placed on top of the core-cladding interface of the waveguide. Therefore, these emitters get illuminated by the mode patterns of the waveguide. Then the final image at the camera plane is the product of the waveguide mode pattern and slowly fluctuating emitters convolved with the PSF (220 nm).

For simplicity, the mode patterns for the straight waveguide considered here are assumed to have a single spatial frequency and oriented only along the vertical-axis as shown in the figure below. Then each image in the image stack is multiplied with a phase-shifted fringe pattern. The phase shift is to mimic for the piezo stage oscillating along the input facet of the waveguide. The phase shift of the interferogram is chosen to change by 2π radians over 10 frames. SOFI$^2$ and SACD$^2$ reconstructions along with the SSIM scores are given below.

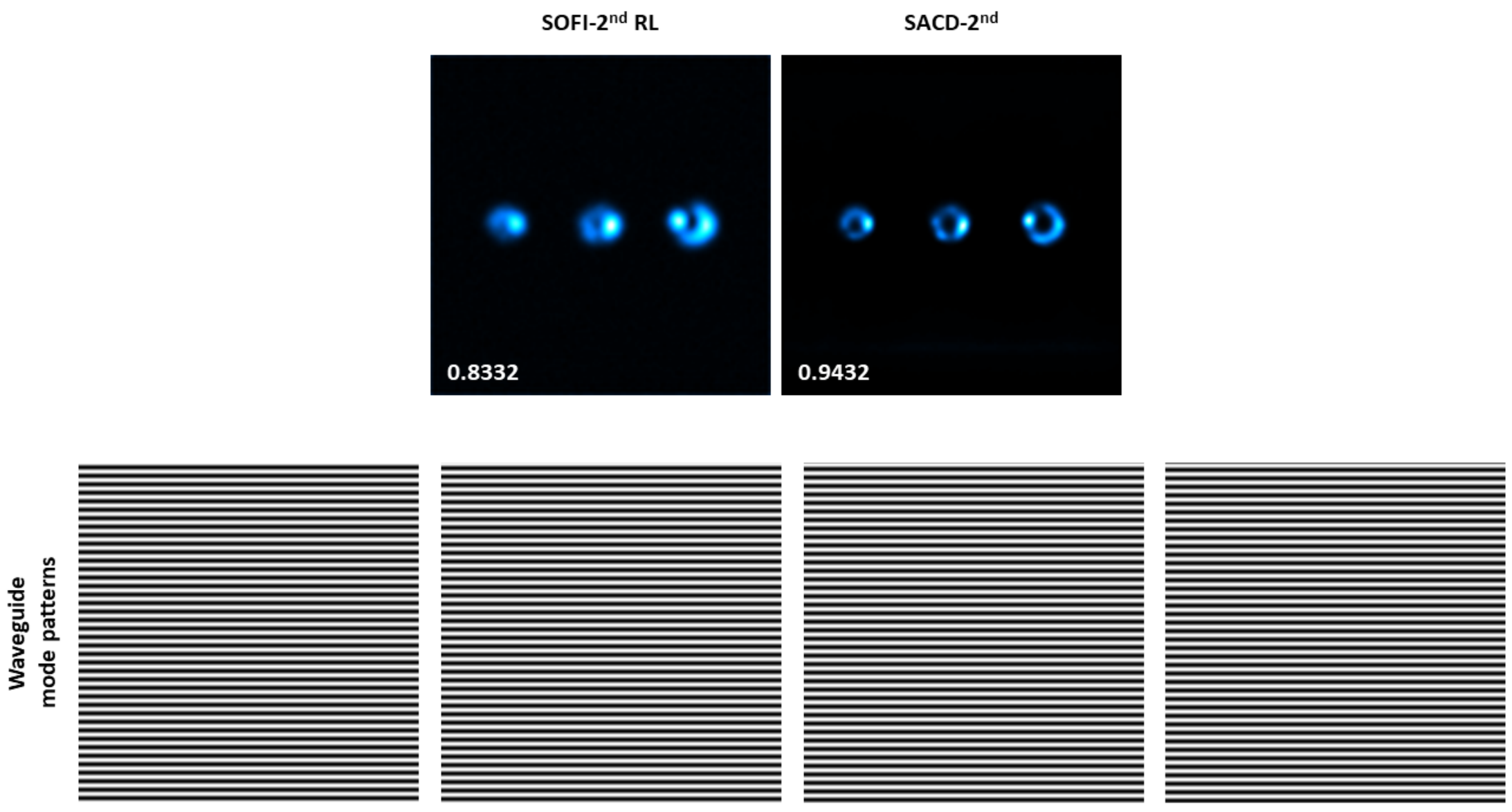

**Figure. 4**: Influence of mode patterns of a straight waveguide on reconstructions using 2nd order SOFI and 2nd order SACD on the synthetic data set with slow intrinsic fluctuations ($\zeta_{on} = 1.67 \times 200$ frames and $\zeta_{off} = 2.5 \times 200$ frames). A few of the phase-shifted mode patterns used for illuminating the sample is also shown.

The SSIM score has improved from 0.2483 to 0.8332 for $SOFI^2$ and from 0.8024 to 0.9432 for $SACD^2$. This implies that the intensity-fluctuations induced by the waveguide mode patterns have helped achieve a better reconstruction. An important observation is that the non-uniform waveguide illumination induces artifacts (unresolved regions) in the image. This can be seen as bright points in the reconstructed synthetic structures in Fig. 4. These points lie along the fringes, i.e., horizontally. This arises due to the correlation between the different emitters.

Simulation3: The experimental particulars used for the emitters are the same as in the previous two cases. Here the difference arises solely because a four-arm waveguide is used to illuminate the slowly fluctuating emitters. The use of a four-arm waveguide helps mitigate the correlation between the emitters due to the random nature of the sub-diffraction sized speckle patterns. This is because these waveguides are highly multi-moded and all the four waveguides overlap in the imaging area as shown in Fig. 2b of the main text. Therefore, as the piezo stage oscillates the coupling objective along the input facet of the waveguide, a different set of modes get excited in the overlapping waveguides and thereby, induces stochasticity in the illumination pattern which helps to mitigate the illumination induced correlation.

A few of the different mode patterns used for illuminating the emitters are provided in the figure below. For simplicity, we have considered mode patterns with a single spatial frequency but with different azimuthal orientations for each frame as shown below in Fig. 5. To introduce the effect of the piezo stage oscillation along the input facet of the waveguide, these interferograms are given a phase shift of $2\pi$ radians over 10 frames. During the experiments, there will be multiple spatial frequencies and the mode patterns will be more chaotic due to the multi-moded waveguides employed.

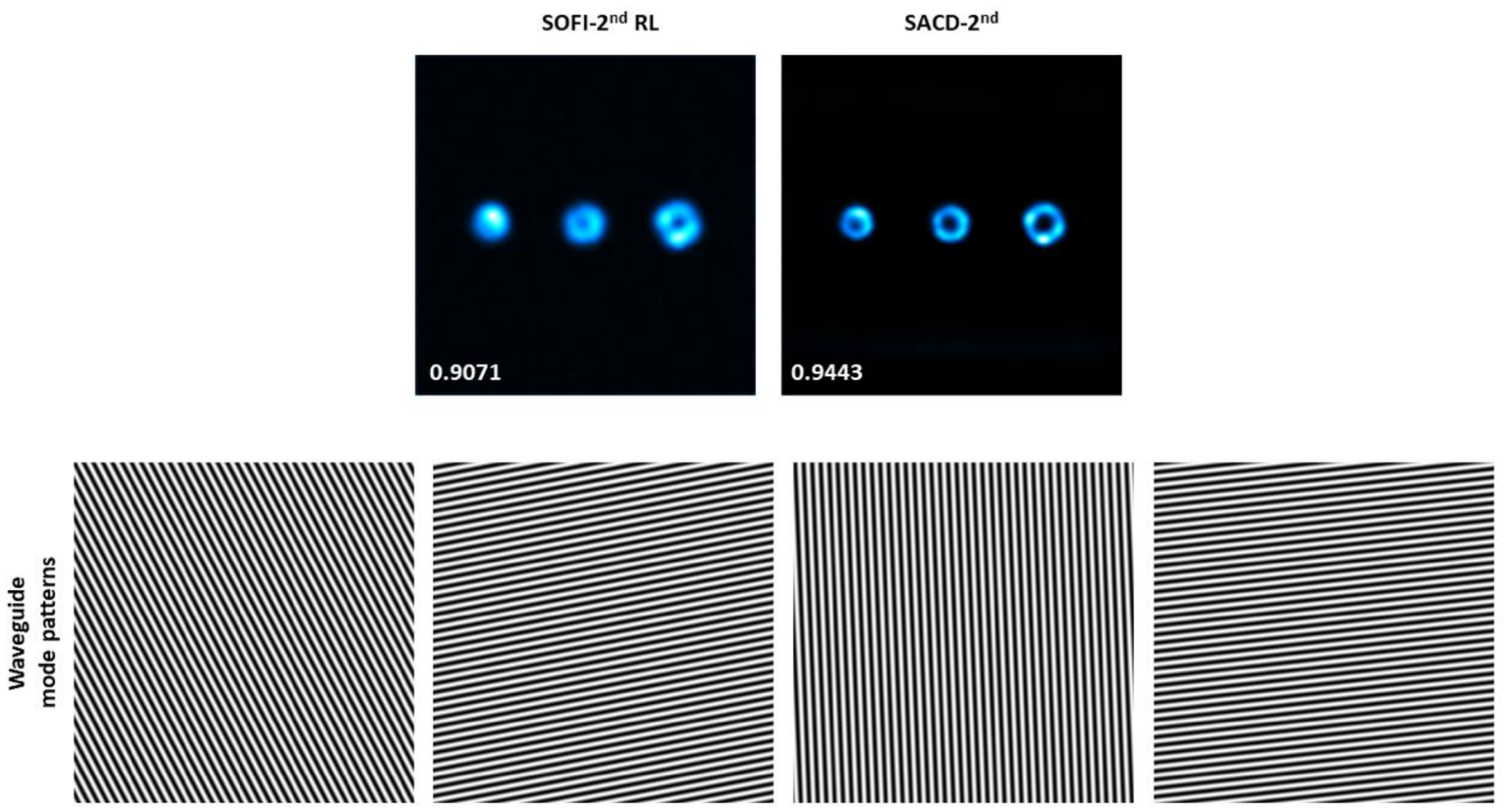

**Figure. 5**: Influence of mode patterns of a four-arm waveguide on reconstructions using 2nd order SOFI and 2nd order SACD on the synthetic data set with slow intrinsic fluctuations ($\zeta_{on} = 1.67 \times 200$ frames and $\zeta_{off} = 2.5 \times 200$ frames). A few of the phase-shifted mode patterns with different azimuthal orientations is also shown.

The main finding is that using a four-arm waveguide, we obtained the highest SSIM score for SOFI$^2$ and SACD$^2$. SSIM score for SOFI$^2$ is 0.9071 and for SACD$^2$ is 0.9443. The artifacts (unresolved areas) have been mitigated due to reduced correlation arising due to different orientations of the illumination patterns in each frame.

Experimental validation: The simulation studies concluded that a four-arm waveguide geometry is ideal for mitigating the correlation value between the different emitters so that SOFI$^2$ and SACD$^2$ can generate super-resolved images. Therefore, in this experiment a four-arm waveguide geometry is employed to illuminate 195 nm polystyrene beads placed on top of its core-cladding interface. An image stack of 100 frames is captured by oscillating the coupling objective mounted on the piezo stage and is given as input to SOFI$^2$ and SACD$^2$. The reconstructed EPSLON images are shown and the correlation values between the emitters at 4 different regions of interest (ROI) are also provided in Fig. 6.

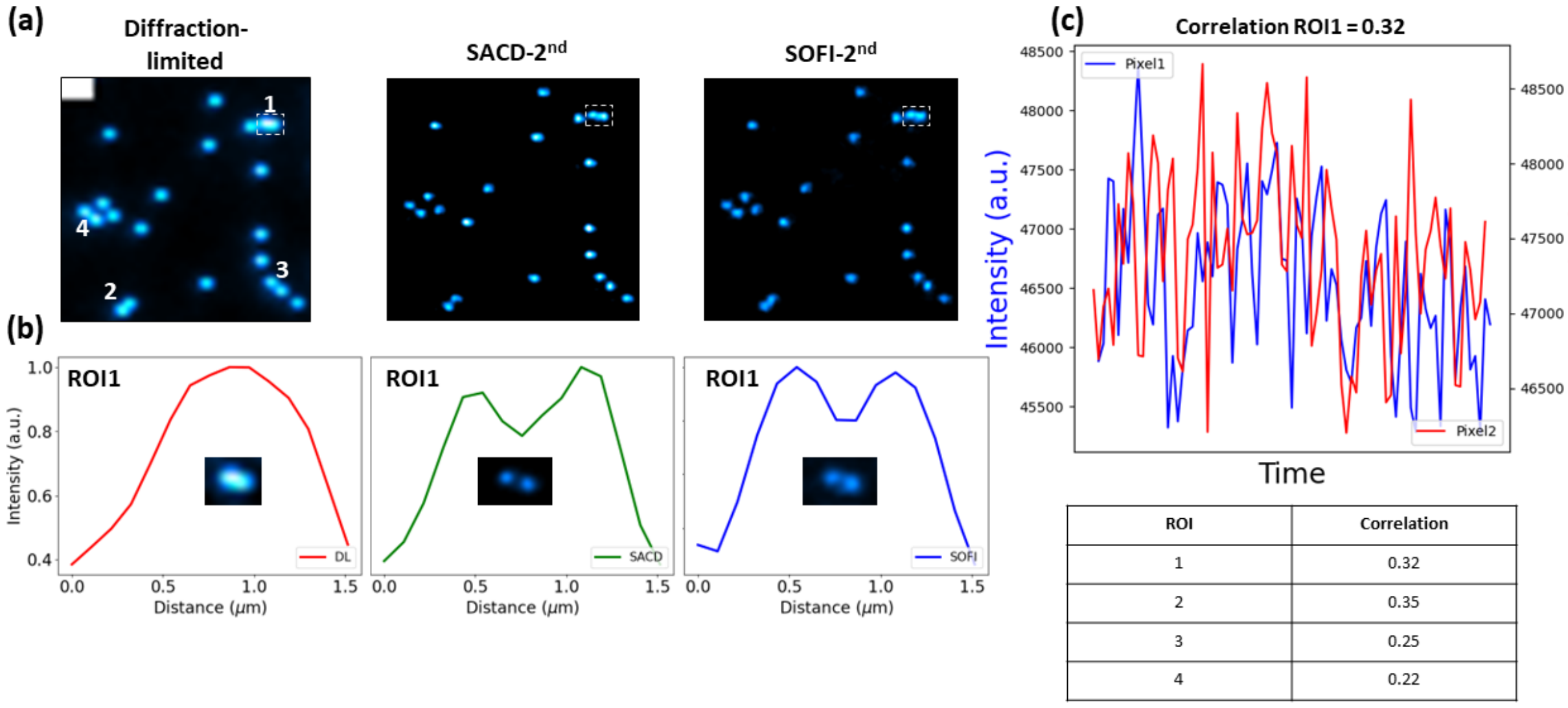

| ROI | Correlation |
|---|---|
| 1 | 0.32 |
| 2 | 0.35 |
| 3 | 0.25 |
| 4 | 0.22 |

**Figure. 6**: **Experimental demonstration of low correlation value in four-arm waveguide geometry for EPSLON**. (a) Diffraction-limited (DL) image and its corresponding $2^{nd}$ order SACD and $2^{nd}$ order SOFI reconstructions are shown, scale bar 2 µm. DL image is the averaged intensity image of 100 frames as the piezo stage oscillates along the input facet of the four-arm waveguide. Four different regions of interest (ROI) labeled '1', '2', '3' and '4' are shown in the DL image. (b) The red, green and blue line plots correspond to the intensity variation (normalized) in ROI1 of DL, $SACD^2$ and $SOFI^2$ images. The insets in the plots provide a magnified view of ROI1 in the DL (unresolved beads) and reconstructed images (super-resolved beads). (c) Line plot shows correlation between the two pixels hosting the unresolved emitters in region '1' in the DL image. The correlation value is 0.32 and the correlation plot is shown. $SACD^2$ and $SOFI^2$ resolve the beads due to the low correlation in the DL image. (d) The correlation values in ROI2, ROI3 and ROI4 are given in the table.

Thus, in our manuscript to avoid the influence of the correlation arising due to active modulation of the PL, two strategies have been adopted:

(i) straight waveguide (Fig. 2a in main text) and $SACD^2$. The high-index core material and interference between the multiple modes in straight waveguide helps generate sub-diffraction sized speckle patterns that is further employed by $SACD^2$.

(ii) four-arm waveguide (Fig. 2b in main text) and $SOFI^2$ or $SACD^2$. The high-index core material plus the counter propagating highly multi-moded waveguides creates sub-diffraction sized speckles that vary stochastically due to the piezo scanning along the input facet of the waveguide. Such a strategy helps generate sub-diffraction sized speckle patterns that can be employed by $SACD^2$, $SOFI^2$ and BlindSIM.

## 4. Schematic diagram of the imaging setup for EPSLON

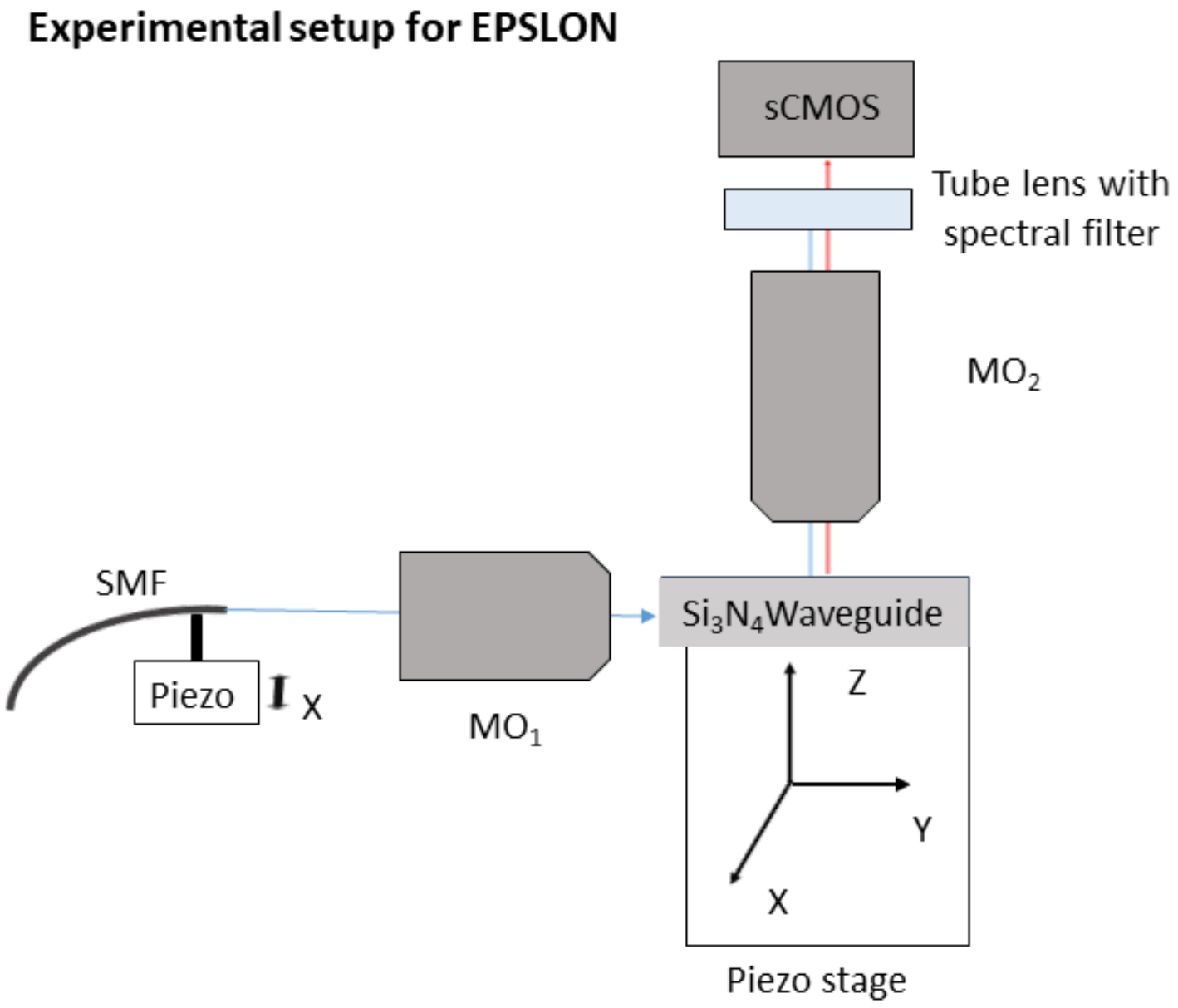

**Figure. 7. Schematic diagram of experimental setup for EPSLON**. Laser guided by a single mode fiber (SMF) held on a XY piezo stage by vacuum chuck is collimated and directed to the back-focal plane of coupling objective Olympus LMPanFL N 50×/0.5 NA $MO_1$. The light is focused by $MO_1$ and coupled into $Si_3N_4$ waveguide mounted on a XYZ piezo stage. The coupled laser light in the waveguide, induces a broadband photoluminescence in the core. Any index perturbation scatters the PL light, shown in red, into the far-field via detection objective MO2, tube lens with spectral filters onto a sCMOS Hamamatsu C13440-20CU camera**.** The spectral filters are chosen to reject the coherent laser light and transmit only the incoherent photoluminescent light.

### Investigation of the PL: Experiment to measure the ratio of transmitted to confined photoluminescence

In the first experiment, the ratio of PL light that is confined inside the waveguide to the PL light transmitted into the far-field is quantified and found to be more than two orders of magnitude. The experimental configuration is shown below in Fig. 8 below. Laser light (488 nm) is coupled into a 400 µm wide $Si_3N_4$ waveguide using a coupling objective. This guided light will induce broadband incoherent photoluminescence (PL) inside the core. PL emitted inside the core will be both transmitted into the far-field and confined inside the core. The confined light will get guided along the length of the waveguide, attenuated by the propagation loss for that wavelength (see Table 1 above). At the output facet of the waveguide, a detection objective 2 is used to collect the guided confined light. A combination of long-pass and band-pass filter ensures that the coherent coupling laser light is blocked and only the incoherent PL guided light reaches the camera. The PL light collected by Detection Objective 2 is shown as inset in Camera which is the experimentally obtained image of the PL emanating from the output facet of the waveguide.

Now, the transmitted at the imaging area is also captured by Detection Objective 1. The same combination of long-pass and band-pass filters ensure that only the transmitted incoherent PL light reaches the camera, and the coherent laser light is blocked. The inset in camera shows the multiple modes overlapping at the imaging area of the waveguide. This experiment is repeated multiple times and the counts reaching the cameras are measured. The ratio of PL transmitted (Detection objective 2) to PL confined (Detection objective 1) is approximately 0.01.

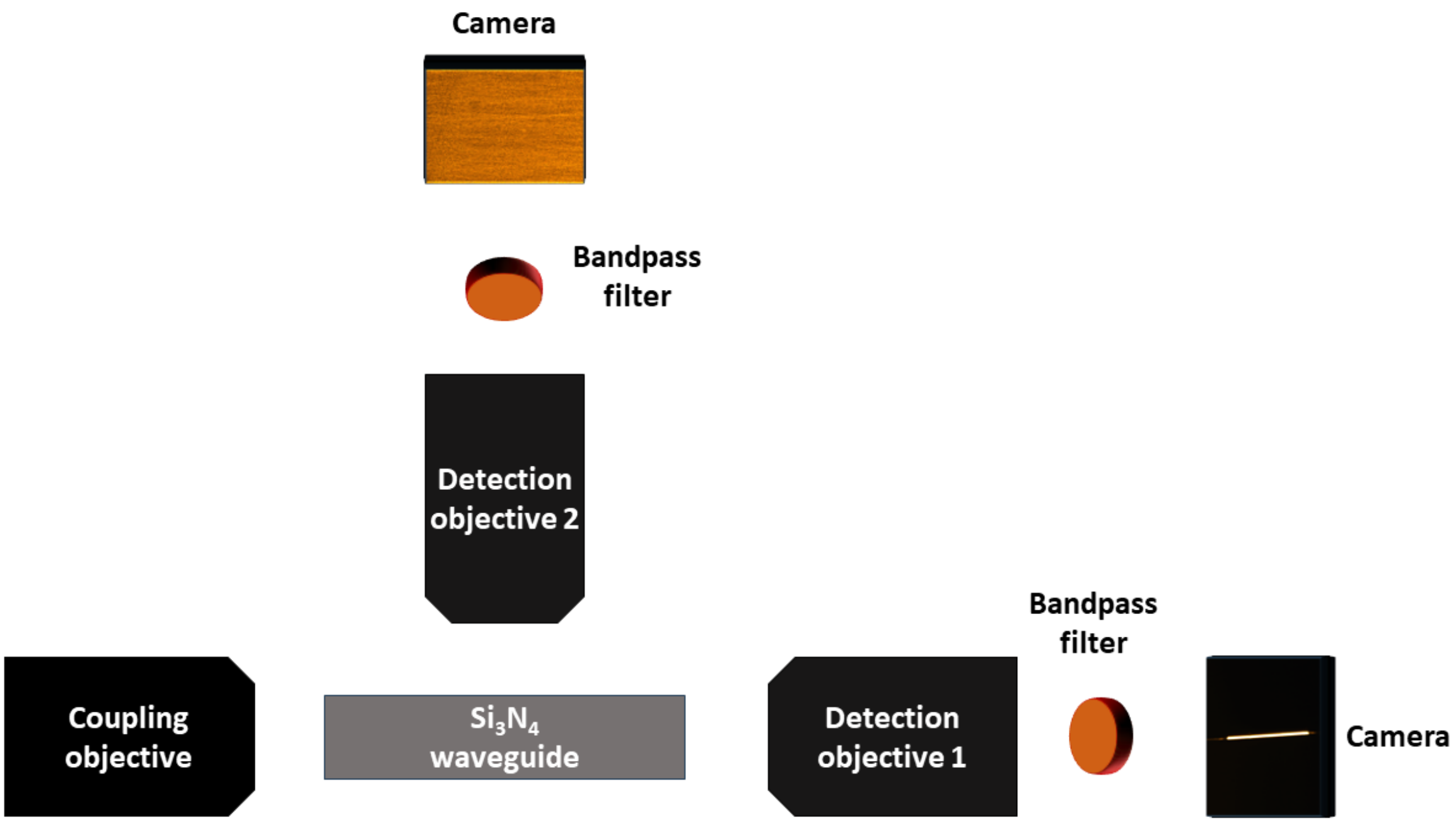

**Figure. 8a: Schematic of the experimental setup to measure the ratio of transmitted photoluminescence to the confined photoluminescence.** A coupling objective is used to guide laser (488 nm vacuum wavelength) along the length of the $Si_3N_4$ waveguide. A combination of long-pass (561 nm EdgeBasic LWP) and bandpass filter (592/43 nm EOTECH SPEC 67034) ensures that the coherent guided light is blocked and only the incoherent photoluminescence reaches the camera. Two identical detection objectives are used: Detection objective 1 for detecting the guided light along the length of the waveguide and Detection objective 2 for detecting the transmitted (unguided light) in the waveguide. Light is detected finally by identical scientific cameras. Actual images detected in the experiment are provided as insets in the cameras. The ratio of light collected by objective 1 and object 2 is found to be more than two orders of magnitude.

## Investigation of the PL: Experimental verification of evanescent nature of EPSLON

This experiment is extended using rat kidney section to validate that it is indeed the evanescently decaying PL light that contributes to the signal. For this experiment, a rat kidney section (embedded in glycerol) is placed on the waveguide. The kidney section lies on top of both $Si_3N_4$-$SiO_2$ layer and in the imaging area ($Si_3N_4$ core) where no $SiO_2$ layer is present. As can be seen in Fig. 8b below, only those portions of the rat kidney section which are placed directly on top of $Si_3N_4$ core-cladding interface scatter light into the camera. The epi-illumination image shows the presence of the tissue section on top of the $SiO_2$ cladding layer. This experimentally proves that the evanescent field of the waveguide core contributes predominantly to image formation in EPSLON.

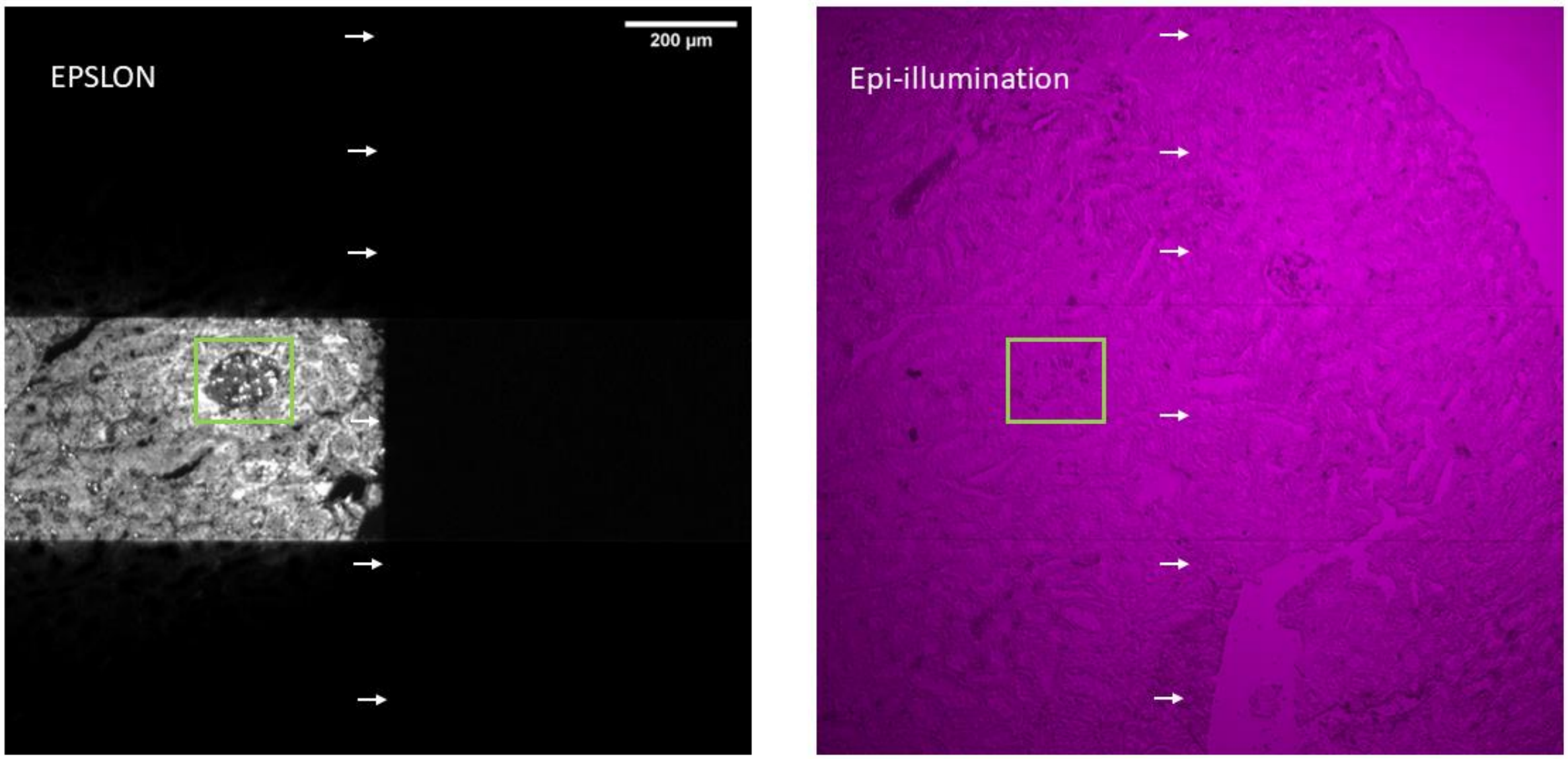

**Figure. 8b: Comparison between EPSLON and epi-illumination.** The green box simply serves as a landmark to correlate between EPSLON and epi-illumination images. The white arrows indicate the boundary where $SiO_2$ cladding layer stops on the waveguide. As can be seen from the EPSLON image, signal reaches the camera from tissue sections lying directly over the $Si_3N_4$ core and no signal is obtained from tissue sections lying over the $SiO_2$ cladding, scale bar 200 µm.

## 5. Optical modes of the waveguide to induce intensity-fluctuations: simulation study

Two identical square particles, 150 nm in size, are placed on top of a 10 µm rectangular waveguide. The center-to-center separation between the particles is 350 nm. The refractive index of the core is set as 2 and refractive index of cladding is set as 1. This waveguide is excited at 500 nm vacuum wavelength. The detection objective numerical aperture is set to NA = 1. Then the coherent transfer function is defined as $h_c = \frac{j_1(\psi)}{\psi}$, where $\psi = \frac{2\pi\, r\, NA}{\lambda}$, where a circular aperture is assumed. Here r is the radial coordinates and λ is the vacuum wavelength of light. For simplicity, emission and detection wavelength is set to be equal, λ=500 nm. Then the spatial cut-off frequency of the system for EPSLON is defined to be twice the abovementioned coherent cut-off frequency, which corresponds to the incoherent case.

The electric field (modes of the waveguide) interacts with these particles. The scattered field is defined as the product of the electric field distribution of the guided mode and the particle. The sample space is defined as an array of zeros except at the location of the two particles, where the value is set to 1. This ensures that only the scattered fields off the two particles propagate into the far-field because in a waveguide-based illumination scheme, only the scattered fields reach the camera plane. Since identical particles are considered, phase difference between the scattered fields arises only due to different locations of the particles as mentioned in the main text. As the piezo stage oscillates the coupling objective along the input facet of the waveguide, different modes get excited with different amplitudes.

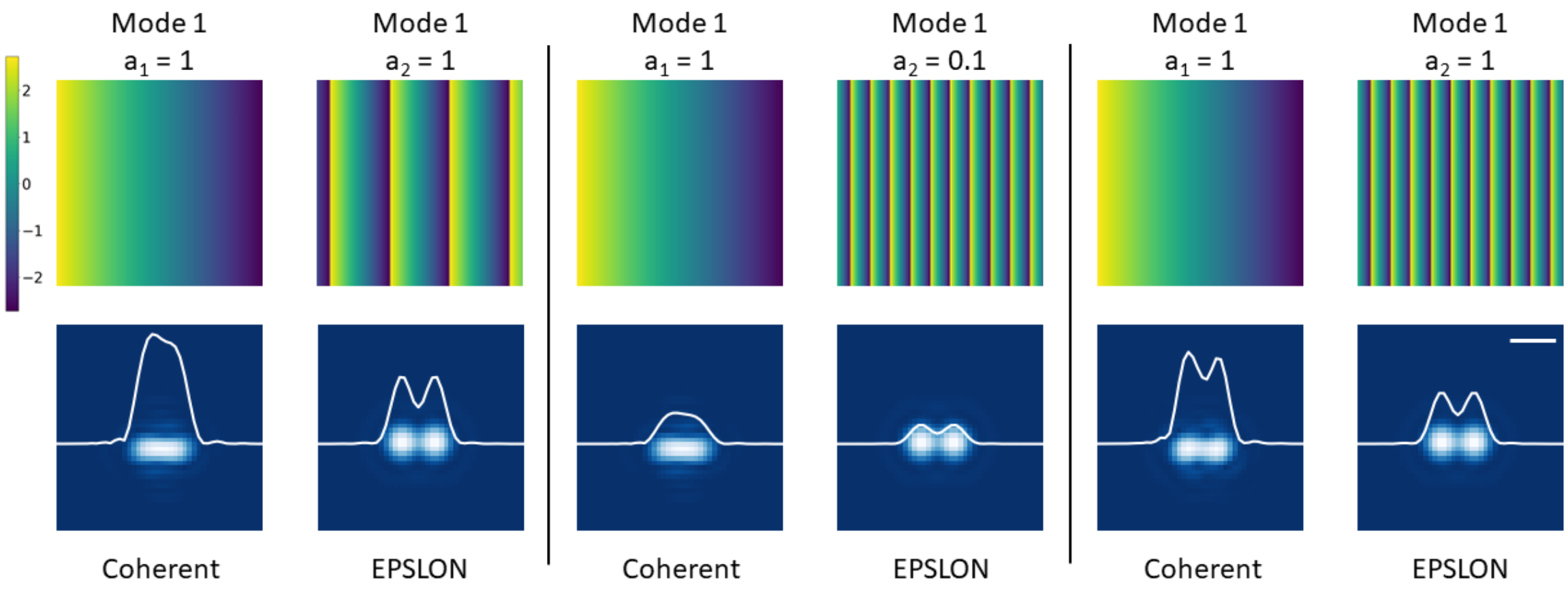

**Figure. 9: Two particle resolution comparison between waveguide-based coherent and incoherent (EPSLON) imaging.** Three separate cases are considered for comparison between coherent and incoherent EPSLON imaging. For brevity, only the fundamental mode with amplitude $a_1$ and higher-order mode with amplitude $a_2$ are considered interacting with the sample in each case. The colorbar provided alongside indicates the phase variation of the field across the cross-section or width of the waveguide. The sample consists of two 150 nm sized particles which are placed with a center-to-center distance of 350 nm apart on the core-cladding interface of a 10 µm $Si_3N_4$ waveguide. The particles scatter 500 nm wavelength light into a detection objective with NA = 1. The image generated at the camera plane for the coherent and EPSLON cases are shown, scale bar 500 nm. The loss in phase information in EPSLON imaging leads to similarity in images for the different excitation cases.

## 6. Structured illumination microscopy using a four-arm junction waveguide

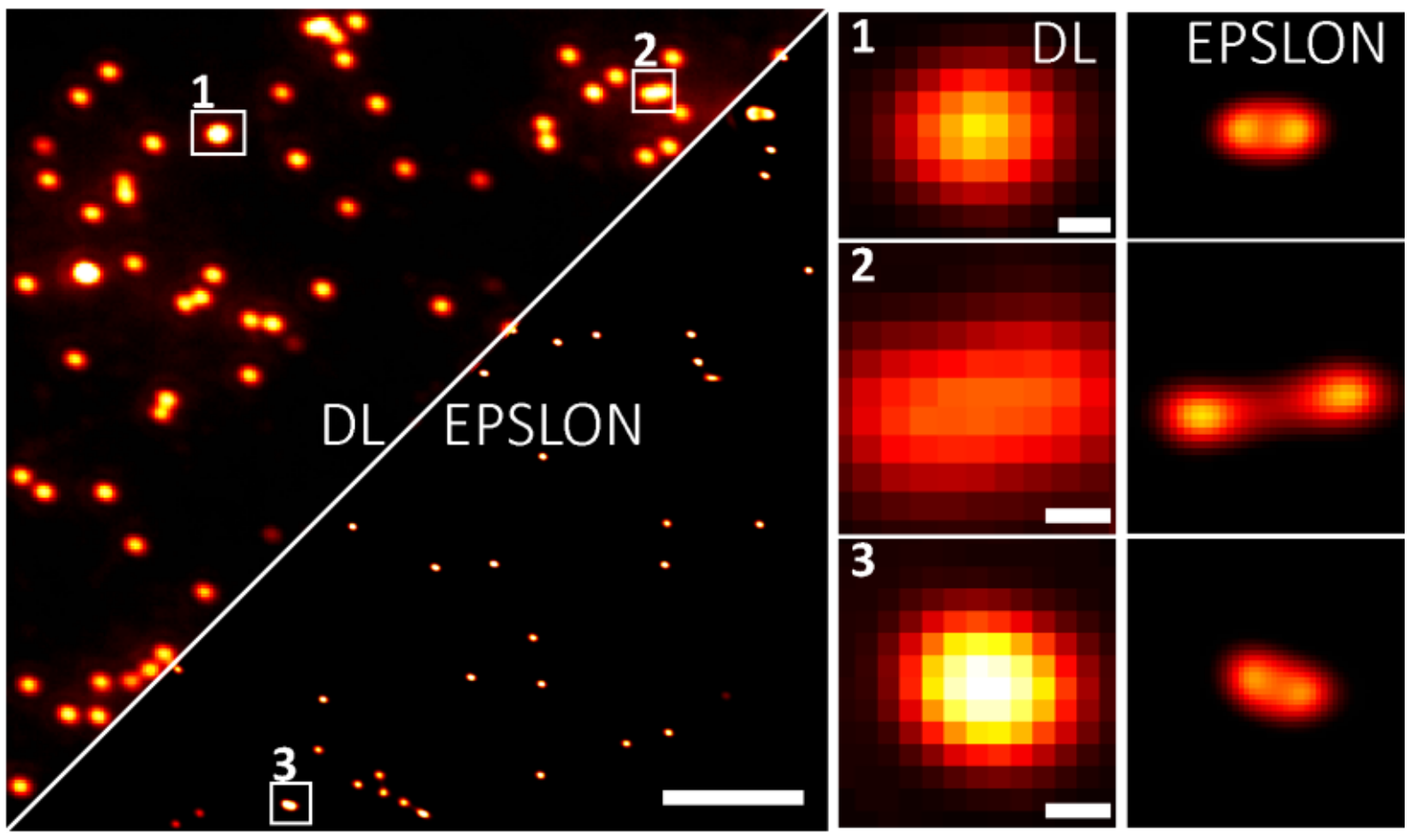

**Figure. 10: Label-free 2D SIM of 195 nm polystyrene beads is demonstrated using 60X/0.9 NA objective**. Diffraction-limited (DL) and super-resolved (EPSLON) images are shown, scale bar 5 µm. Three regions of interest labelled '1', '2' and '3' in DL and their corresponding EPSLON images are blown up and shown alongside, scale bar 125 nm.

## 7. One-dimensional structured illumination microscopy using a SIM chip

For the 1-D SIM experiment, images are acquired using a detection MO with NA = 1.2. Phase-shifted frames required for the SIM reconstruction is generated by temporarily changing the index on one of the arms of the interfering waveguides. The three phase-shifted frames are then given as input to the Fiji plugin of FairSIM. The reconstructed images and its Fourier spectra are provided in Fig. 11 below. The SIM reconstruction can clearly resolve the beads enclosed in the red inset in the DL image. The beads in the green inset in the EPSLON image are separated by 274 nm, shown in the line profile. Due to aberrations in the system, the experimental diffraction-limit of the SIM microscope is quantified to be 500 nm while the theoretical Abbe diffraction limit is $\frac{\lambda_{det}}{2NA_{det}} \approx 272 - 287\ nm$.

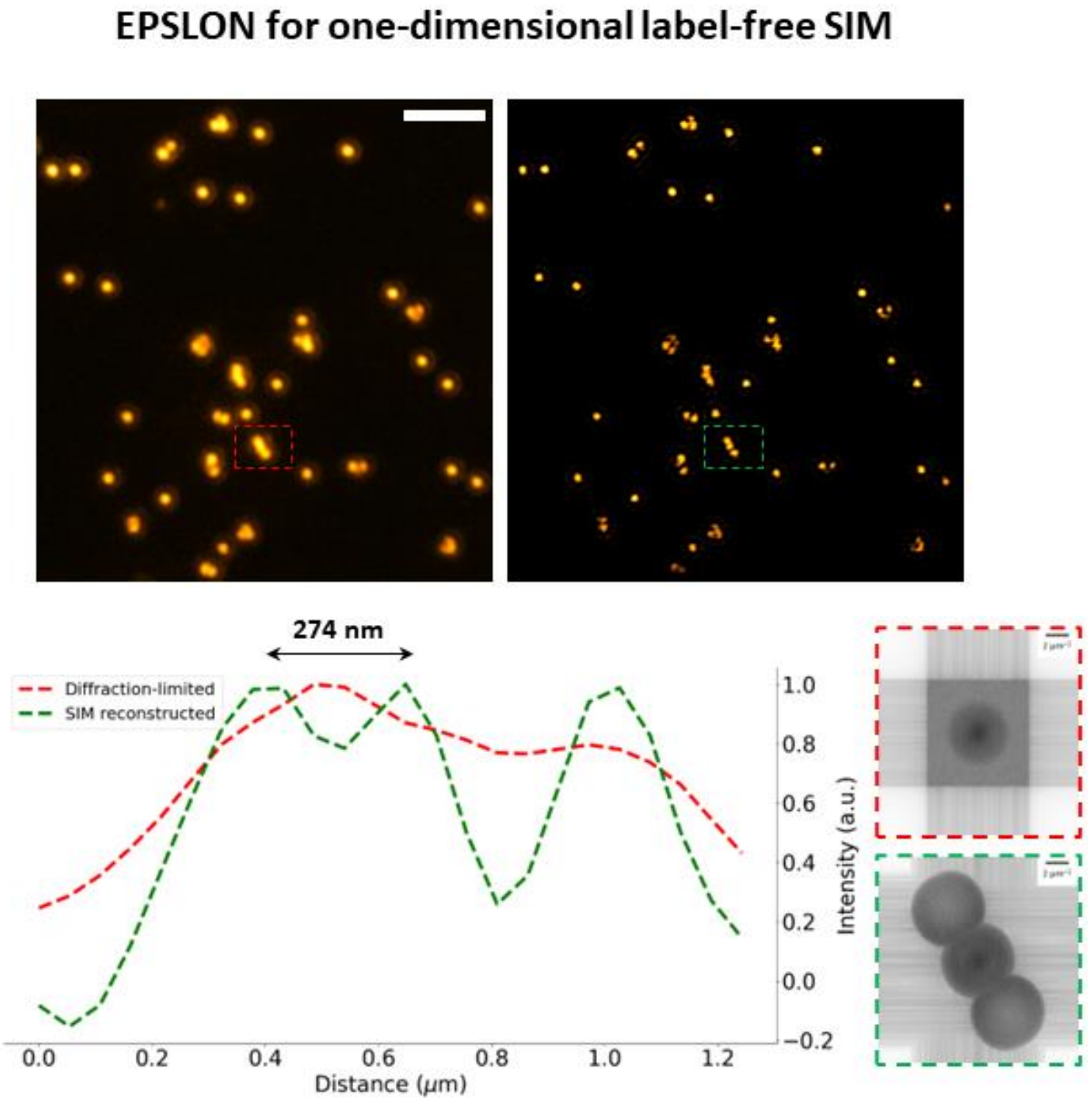

**Figure. 11:** Label-free 1-D SIM of 200 nm gold nanoparticles is demonstrated, scale bar 5 μm. The line profile indicates the intensity variations in the red and green insets in the diffraction-limited DL and FairSIM reconstructed EPSLON images respectively. The line profile clearly shows the separation of two particles spaced 274 nm, which is beyond the diffraction-limit of the imaging system. The Fourier domain representation of the diffraction-limited and reconstructed image are also provided alongside, scale bar 2 $\mu m^{-1}$.

## 8. Investigation of the PL: Speckle size determination

A structured illumination microscopy (SIM) waveguide chip and modes are excited at 640 nm vacuum wavelength (coupled light). The period $f$ of the fringes generated is given by

$$f = \frac{\lambda_{ex}}{2n_f \sin\frac{\theta}{2}}$$

where $\lambda_{ex}$ is the excitation wavelength, $n_f \approx 1.7$ is the refractive index of the guided mode for the $Si_3N_4$ waveguide used here and $\theta$ is the angle between the interfering waveguides.

First, we experimentally measure the fringe period generated in photoluminescence (PL) configuration. The angle between the interfering waveguides in the SIM chip is ≈25°. The period of the fringes is show in Fig. 12 below in the line plot.

Then we coat the imaging area with a very thin layer of cell mask deep red fluorescent stain and excite the waveguide at the same wavelength 640 nm, i.e., acquire a TIRF image of a layer of fluorescent stain. The period of the fringes generated in fluorescence mode (TIRF) is also shown in Fig. 1 below.

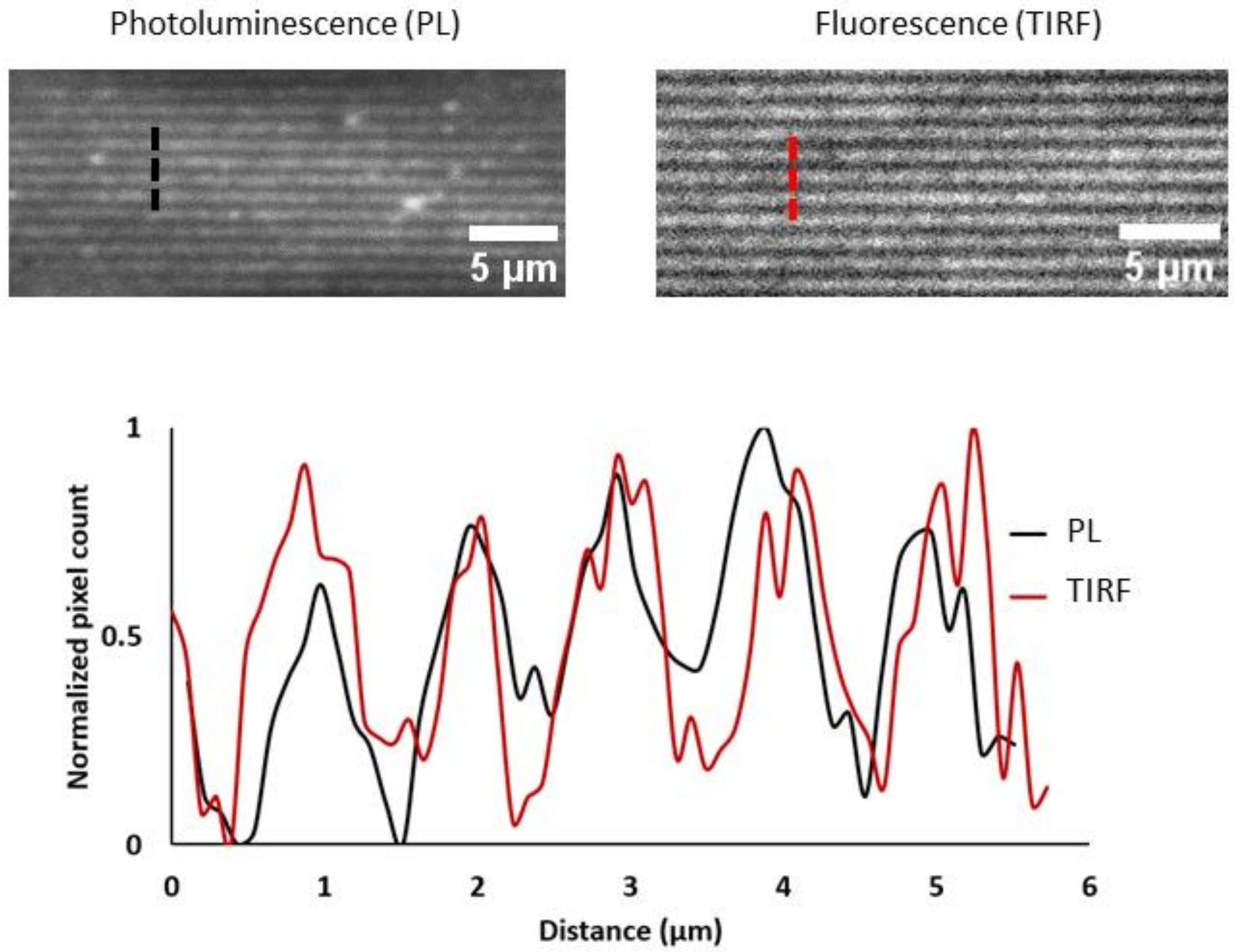

**Figure. 12**: Periodicity of fringe patterns in PL and TIRF mode using a ≈25° SIM chip when $\lambda_{ex} = 640\ nm$. The line plots show the normalized modulation in intensity across the black and red dotted lines in the PL and TIRF images respectively. White scale bar 5 µm.

The period of the fringes (≈ 870 nm) match in both PL and fluorescence mode and closely match with the theoretical value predicted by the equation above. This implies that period generated by the SIM chip in PL and TIRF mode is dependent only on the excitation wavelength $\lambda_{ex}$, index of the guided mode and angle between the interfering waveguides as mentioned in the formula above. The period in the SIM chips is not dependent on the emission autofluorescence wavelength $\lambda_{af}$.

Next, to show fringe period dependence on the excitation wavelength dependence $\lambda_{ex}$, the same SIM chip used above is excited in PL configuration at 640 nm and 561 nm. The periodicity and contrast of the fringe patterns are also provided alongside in Fig. 13.

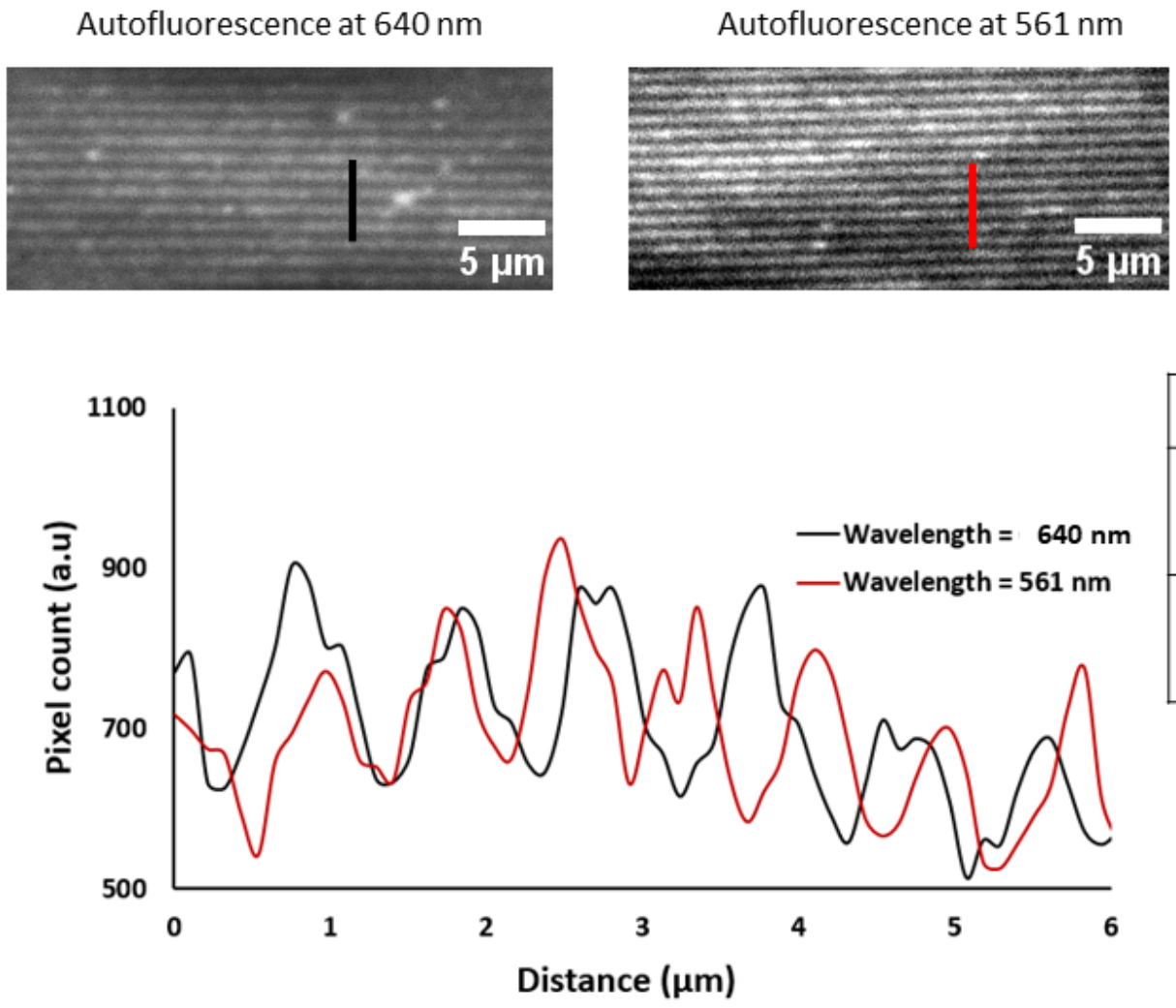

| Sl. No. | Mode | Period | Contrast |
|---|---|---|---|
| 1 | Autofluorescence at 640 nm | ≈ 870 nm | ≈ 0.14 |
| 2 | Autofluorescence at 561 nm | ≈ 756 nm | ≈ 0.15 |

**Figure. 13**: Fringe period in PL mode using a ≈25° SIM chip when $\lambda_{ex} = 640\ nm$ and $\lambda_{ex} = 561\ nm$. The line plot shows the modulation in intensity across the black and red lines in the fringe patterns. Black curve and red curve show the modulation in intensity in PL mode at $\lambda_{ex} = 640\ nm$ and $\lambda_{ex} = 561\ nm$ respectively. The table shown alongside mentions the fringe period and its corresponding contrast at $\lambda_{ex} = 640\ nm$ and $\lambda_{ex} = 561\ nm$. White scale bar 5 µm.

From Fig. 13, it is seen that the fringe period scales with $\lambda_{ex}$ as mentioned by the formula earlier. The implication of the experimental results in Fig. 12 and Fig. 13 shown above are that, for a fixed interference angle between the chips, fringe period is determined by $\lambda_{ex}$ and its corresponding effective mode index. The fringe period is ≈ 870 nm and contrast is ≈ 0.14 at $\lambda_{ex} = 640\ nm$. At $\lambda_{ex} = 561\ nm$, the fringe period is ≈ 756 nm and contrast is ≈ 0.15 and these periods closely match with the theoretical value given by the formula.

To verify that indeed we have a single spatial frequency corresponding to the excitation wavelength $\lambda_{ex}$, Fig. 14 is provided. As can be seen from the Fourier domain representation of the fringe pattern, the first order peaks, encircled within the red and green regions in Fig. 14, have a single spatial frequency component. Therefore, the fringe period in these SIM chips correspond to the excitation wavelength (monochromatic coupling laser light) and not with multiple frequencies corresponding to the polychromatic emission of PL.

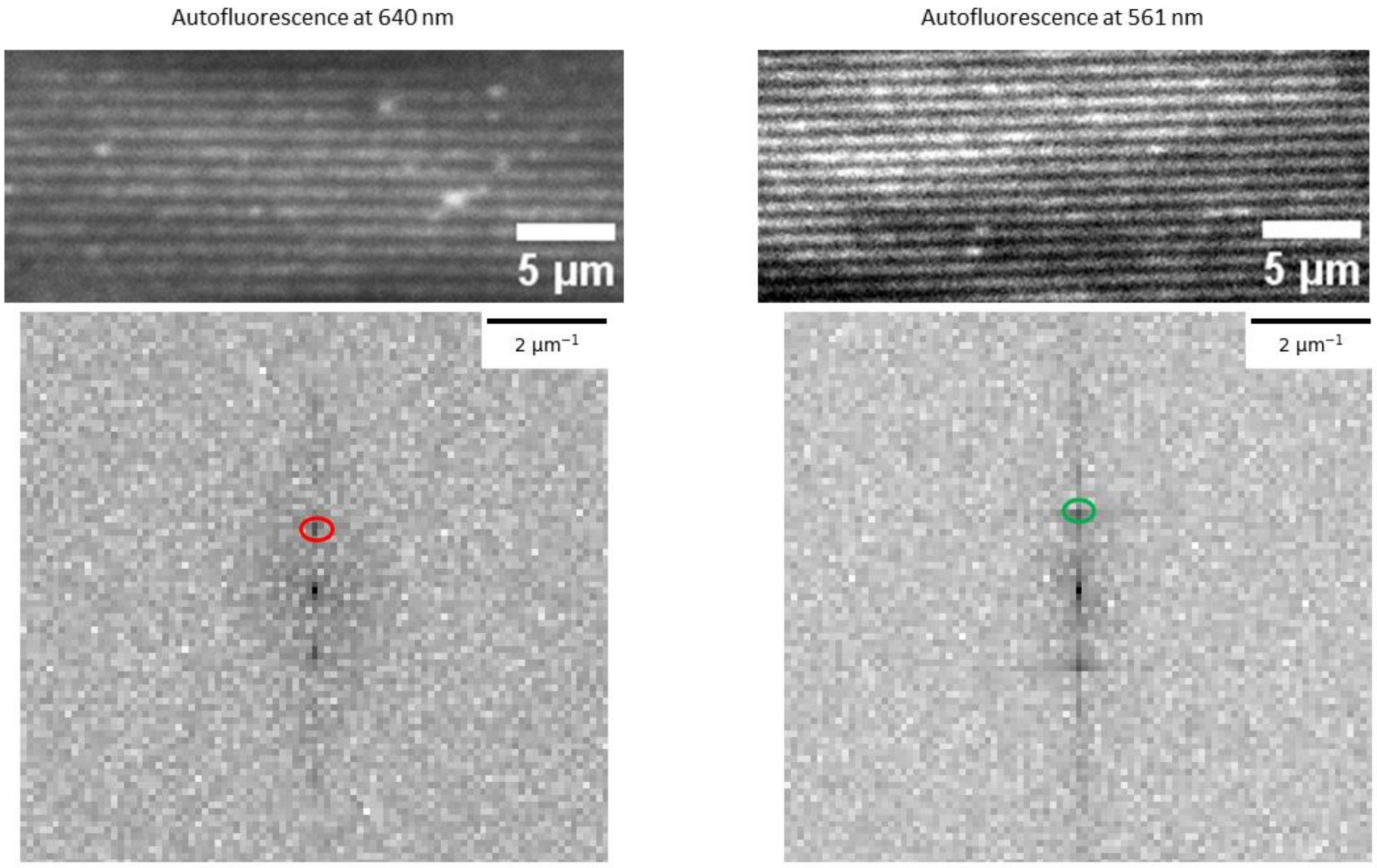

**Figure. 14**: Fringe period in PL mode using a ≈25° SIM chip when $\lambda_{ex} = 640\ nm$ and $\lambda_{ex} = 561\ nm$, scale bar 5 µm. The corresponding Fourier spectrum is shown. The red and green circle indicates the first order component which corresponds to $\lambda_{ex} = 640\ nm$ and $\lambda_{ex} = 561\ nm$. Scale bar 2 $\mu m^{-1}$.

Finally, to demonstrate the influence of detection wavelength $\lambda_{det}$, the following experiment is carried out and the results are shown in Fig. 15. As shown in Fig. 2(e) of the main paper, the PL emission spectrum is very broad spanning more than a few hundred nanometers. Therefore, if we excite the waveguide at 488 nm, we can choose any of the filters, (FITC, TRITC, CY5), provided in Table 2 in the supplementary section. The choice of filters during experiments in this manuscript was based on maximizing the signal at the camera plane, so that the fluorescence-based super-resolution algorithms generated super-resolved images with fidelity.

To experimentally demonstrate that the influence in changing the emission filter (FITC, TRITC, CY5) is in the resolution of the final diffraction-limited image, a straight waveguide is excited at 488 nm. The scattering image and the PL images in FITC, TRITC and CY5 channels are shown alongside. As seen, the speckle patterns in all the images match well. The resolution in each channel is computed using Fourier ring correlation (FRC) and shown in the table alongside.

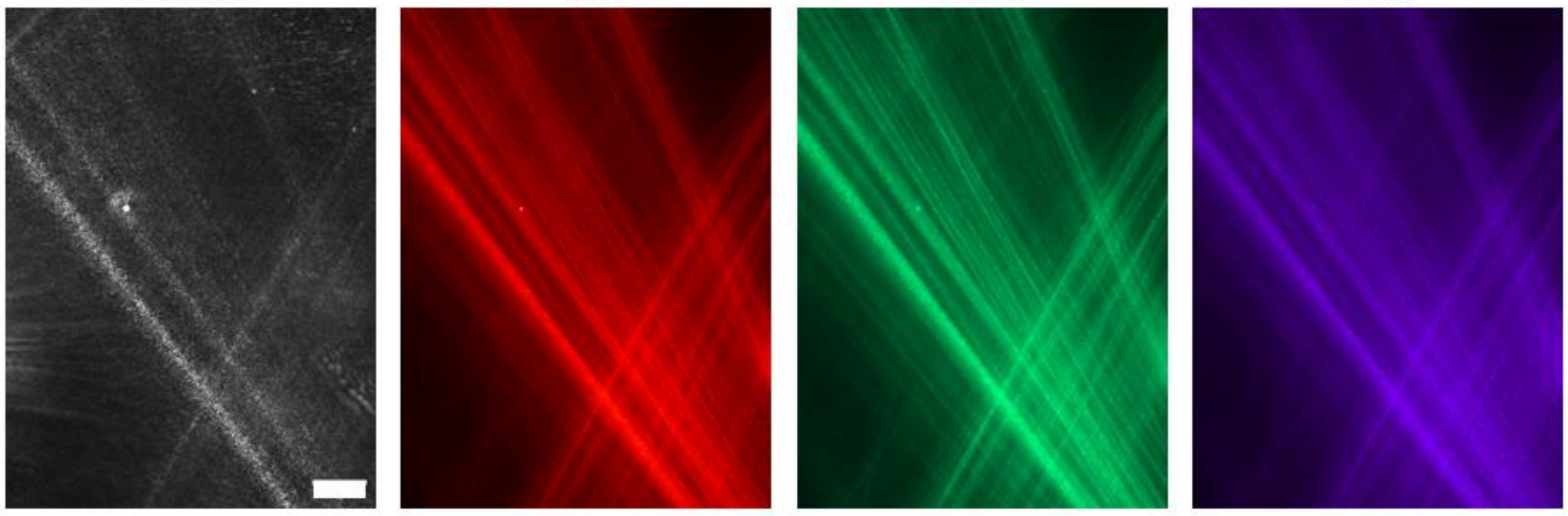

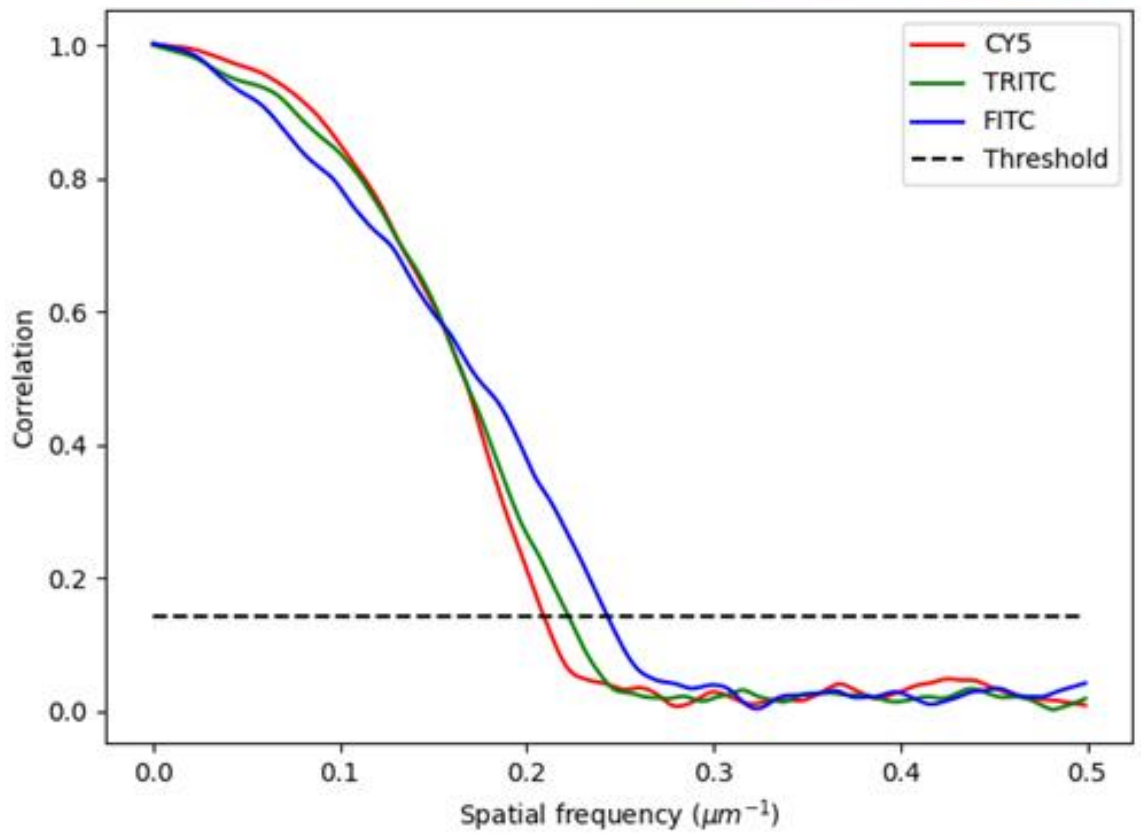

| Sl. No. | Filter | FRC resolution (nm) |
|---|---|---|
| 1 | CY5 (655 nm LP + 690/50 nm BP) | 516 |
| 2 | TRITC (585 nm LP + 630/75 nm BP) | 485 |
| 3 | FITC (515 nm LP+535/30 nm BP) | 444 |

**Figure. 15**: Speckle patterns, in a straight waveguide excited at 488 nm, in different filter channels and the corresponding resolution in each of the channels computed using Fourier Ring Correlation. Scale bar 25 µm.

## 9. Small EV preparation and characterization

Small EVs were isolated from fresh urine samples collected in the morning from healthy donors. The collection of urine samples was approved by the Norwegian Regional Committees for Medical and Health Research Ethics and the participants gave informed written consent. Small EVs were isolated by differential centrifugation as previously described [2]. Briefly, urine (around 200 ml) was centrifuged at 2000×g for 15 min at room temperature (RT) to remove cells and cell debris, and then at 10,000×g for 30 min at RT to separate large particles/vesicles. The resulting supernatant was centrifuged at 100,000×g for 70 min at RT in a Ti70 fixed-angle rotor (Beckmann Coulter, IN, USA) to pellet small particles. The pellet was washed with 20 ml phosphate-buffered saline (PBS) and centrifuged again at 100,000×g for 70 min at 4°C in a Ti70 rotor. The pellet was then resuspended in 6.5 ml PBS, vortexed and centrifuged at 100,000×g for 70 min at 4°C in an MLA-80 fixed-angle rotor (Beckmann Coulter, IN, USA). The supernatant was then removed, and the pellet resuspended in 200 µl PBS (filtered through a 0.02-µm Anotop 25 filter) and stained with CellMask™ Deep Red plasma membrane dye (C10046, Invitrogen, MA, USA) according to manufacturer's instructions. Briefly, small EVs were incubated with CellMask™ Deep Red (diluted 1:500) for 10 min at 37°C, then the unbounded dye was removed and stained EVs washed with filtered PBS using ultrafiltration devices (Amicon Ultra 0.5 mL - 3K, UFC5003234, Millipore, MA, USA) at RT. The sample was then stored at 4°C until further use. A small aliquot of the sample was used to measure the size and number of particles in the 100,000×g pellets using a Nanosight NS500 instrument (Malvern Panalytical, Malvern, UK). The sample was diluted to the optimal working concentration of the instrument ($2 \times 10^8$ to $1 \times 10^9$ particles per ml) with filtered PBS, and then measured. Five videos of 60 sec were acquired and subsequently analyzed with the NTA 3.4 software, which identifies and tracks the center of each particle under Brownian motion to measure the average distance the particles move on a frame-by-frame basis. As shown in Fig. 16, the majority of the small EVs has a diameter between 100-175 nm (65,9% of the total) with a mode of 101 nm.

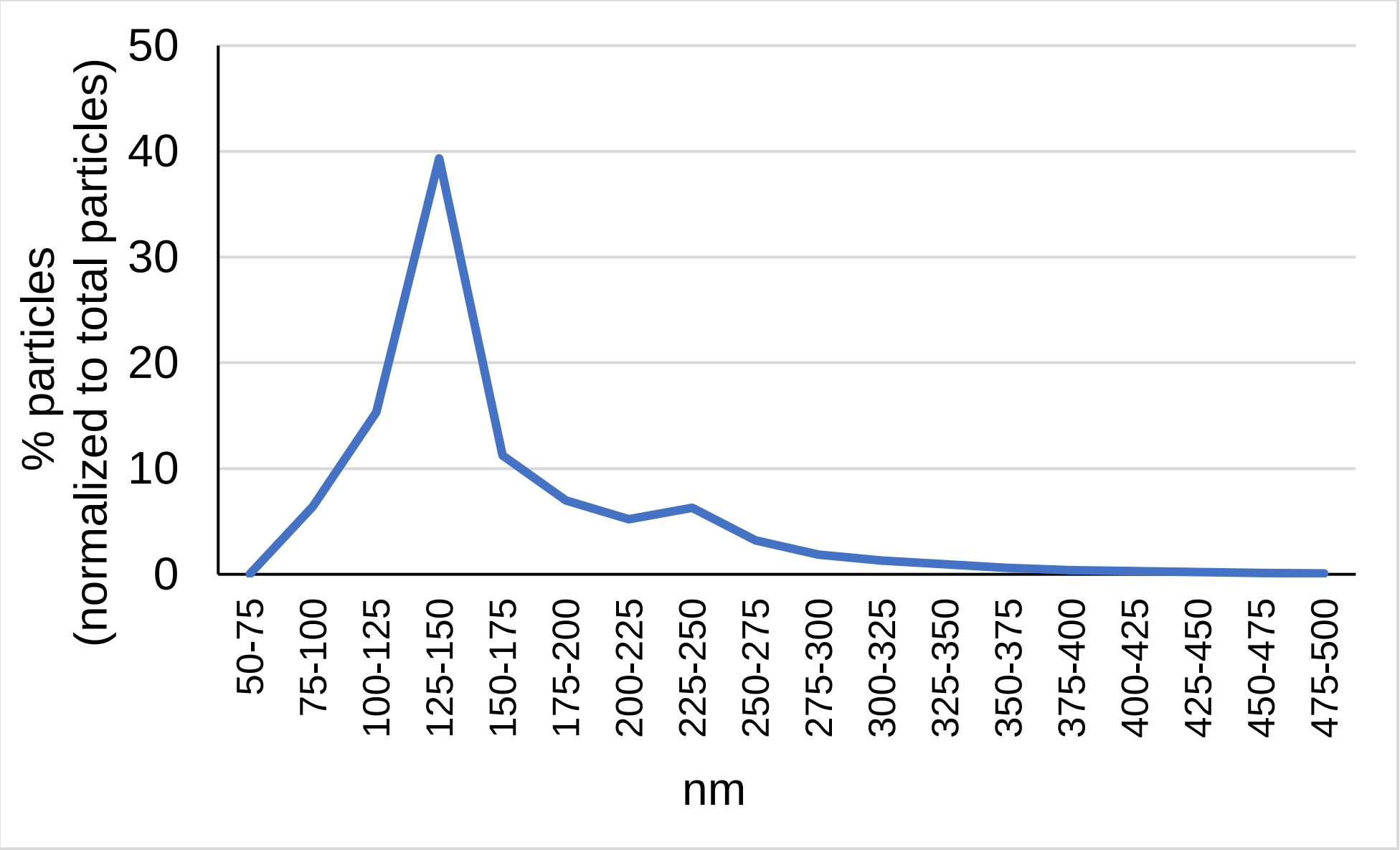

**Figure. 16:** Small EVs were isolated by sequential centrifugation from healthy donor urine and their size was measured by NTA. The size distribution of small EVs is shown as percentage of particles having the indicated size normalized by the total number of particles.

## 10. Rat kidney sections

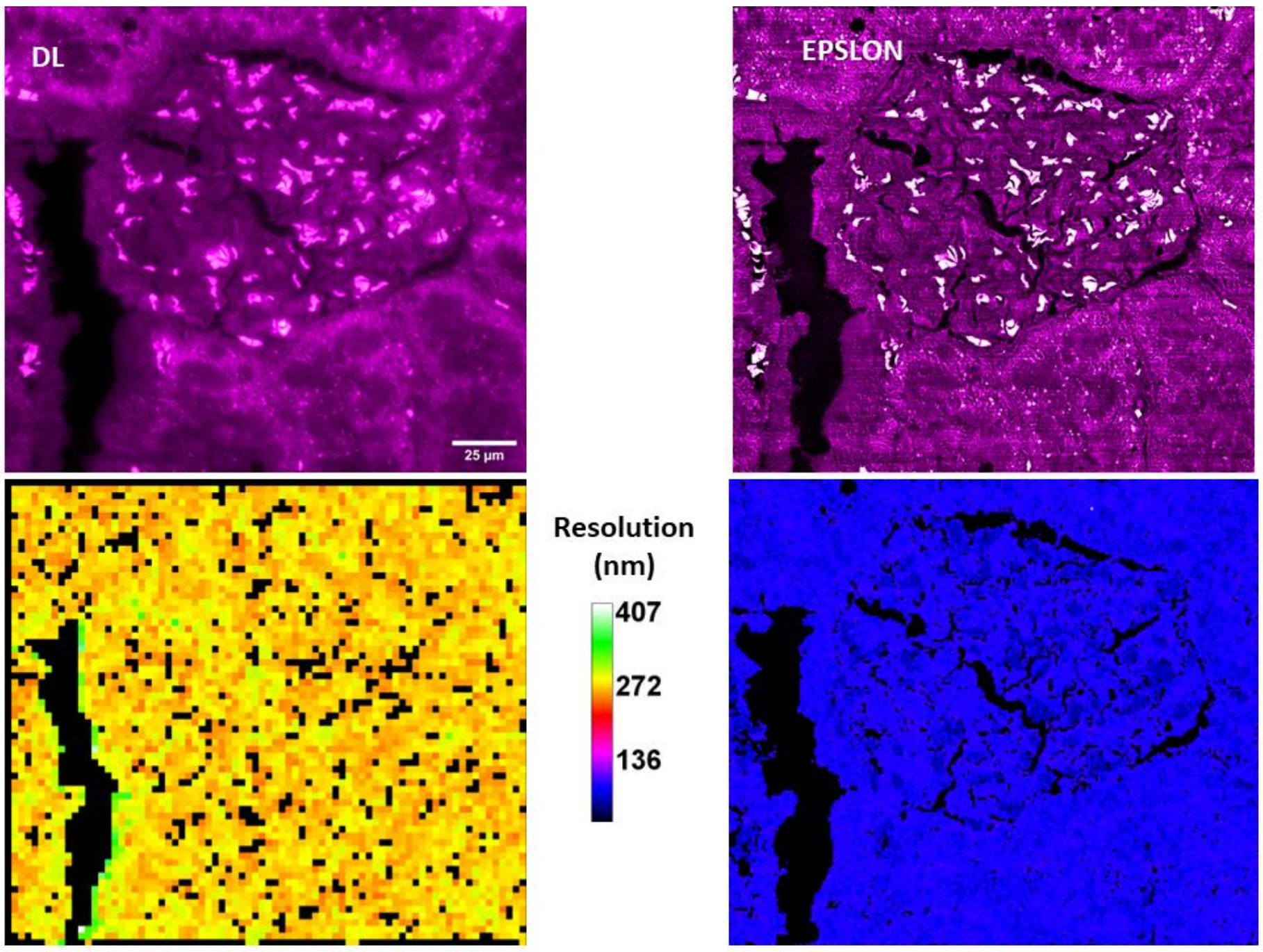

**Figure. 17: Local FRC plot of rat kidney sections shown in Fig. 6 in main text.** Label-free diffraction-limited (DL) and its corresponding super-resolved EPSLON images are shown, scale bar 25 µm. Local FRC resolution is computed and shown below for both DL and EPSLON images. The colorbar indicates the spatial resolution in nanometers.

## 12. Human kidney sections

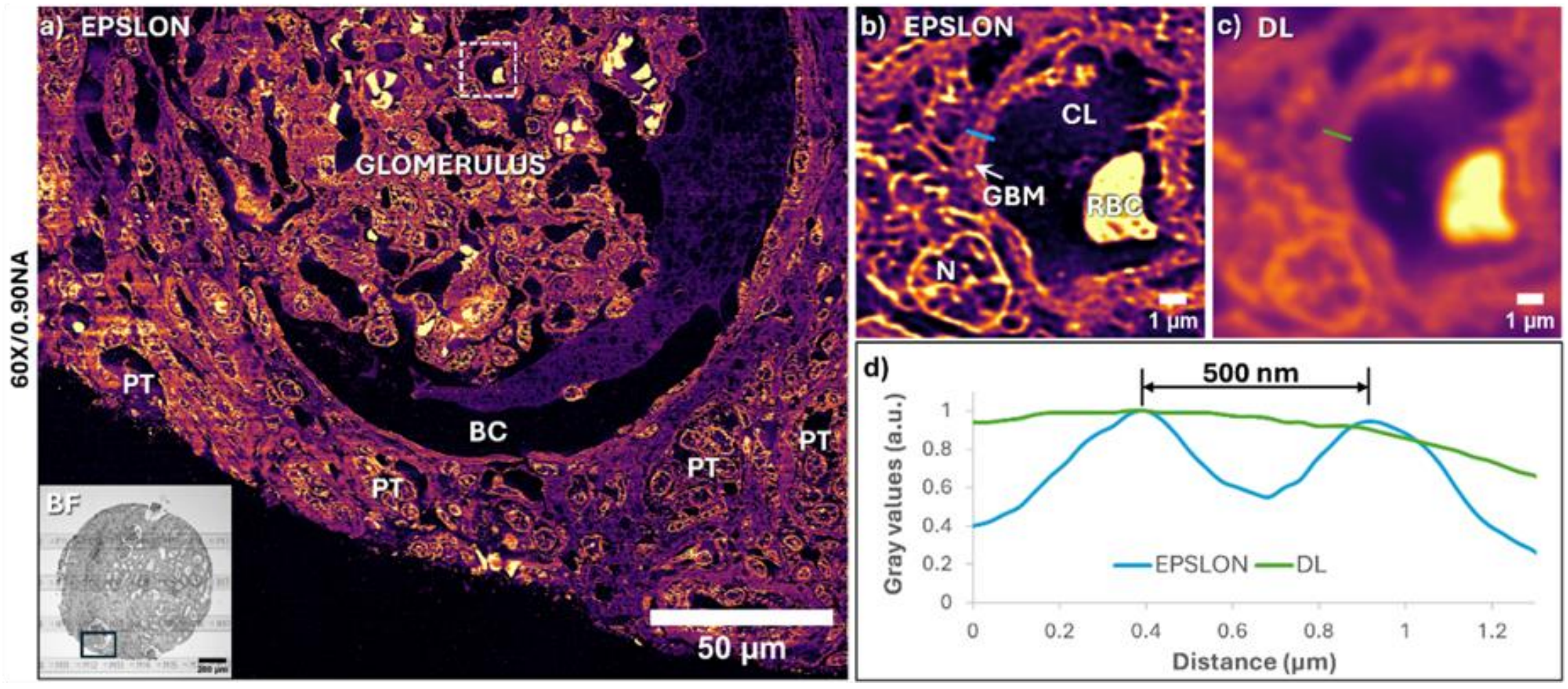

**Figure 18: EPSLON via SACD for human kidney histopathology. (a)** Lable-free super-resolved EPSLON image of a human kidney tissue section is shown, scale bar 50 µm. A mean FRC resolution of 144 nm is observed. The gray scale inset is the bright field image of the kidney section imaged with a 4X/0.10 NA detection objective, scale bar 200 µm. The pseudo-colour image illustrates the EPSLON results of the selected glomerular region in the BF image enclosed within the black box. EPSLON provides a high-contrast contextual visualization of the sample, enabling the identification of structures such as the glomerulus, the Bowman's capsule (BC), and many adjacent proximal tubuli (PT). **(b)** The region in the EPSLON image enclosed within the white dotted box is blown-up and shown, scale bar 1µm. This figure reveals microanatomical structures including a glomerular basement membrane (GBM), a capillary lumen (CL) with a red blood cell (RBC), and the nucleus (N) of an adjacent podocyte. **(c)** The corresponding diffraction-limited (DL) view of the FOV in (b) is shown, scale bar 1µm. Note the poor contrast and resolution as compared to the EPSLON image. **(d)** Line profile measurements over the GBM reveal a separation of approximately 500 nm using EPSLON, which otherwise cannot be measured in DL modality. The blue line represents intensity variation in the EPSLON image along the region shown in (b) and green line represents intensity variation along the DL region shown in (c).

## 13.Human-placenta tissue label-free imaging, preparation and characterization

We now showcase the potential of EPSLON for the evaluation of human placental tissue sections using straight waveguides in tandem with SACD. The human placenta is a pregnancy-specific organ that plays a key role in mediating the gas and nutrient exchange between the mother and the fetus. Moreover, scientific evidence suggests a relationship between ultrastructural changes in placentas and pregnancy-related diseases such as preeclampsia [3].

Placental tissue samples are first fixed in formalin and then embedded in paraffin [4]. 4 μm sections of the tissue samples are then cut from these paraffin blocks using microtome (HM 355S Automatic Microtome, Thermo Fisher Scientific, Waltham, Massachusetts, USA). The cut sections are then placed on poly-L-lysine coated $Si_3N_4$ waveguide chips, and deparaffinized in xylene (3 × 5 min), followed by rehydration in descendent series of ethanol: 100% (2 × 10 min), 96% (2 × 10 min) and 70% (10 min).

Traditionally, the visualization of such changes was made possible through complex, costly, and slow methods such as electron microscopy. Although recent advances in fluorescence-based super-resolution microscopy have successfully enabled detailed visualizations of such ultrastructural features [5, 6], the fluorescent markers used in these imaging methods exhibit practical limitations for routine histological practice. These include (1) low stability, in the sense that the fluorescence imaging must be carried out in a relatively short period of time after sample labelling, e.g. less than a week, to avoid the issues with the decay of the fluorescent dyes; (2) special handling of the fluorescently-labeled sample, particularly to avoid light exposure that might render the dyes photobleached; and (3) extended sample preparation, in terms of cost and time, associated with the fluorescence labelling. Thus, a label-free method for achieving ultrastructural visualizations of the placental morphology would prove advantageous for the field of placental research. Here, the proposed EPSLON method alleviates these issues, enabling a fast, simple, and repeatable route for observing, for example, sub-diffraction sized placental features. We illustrate the potential of EPSLON for placental histology in Fig. 18. First, 250 optical images of a region of interest were acquired using a detection MO with NA = 1.2, and subsequently averaged to obtain a diffraction-limited image (DL), as shown in Fig. 5a. The corresponding super-resolved EPSLON image is also shown in Fig. 18a. Next, to serve as a ground truth, a scanning electron microscope image of the same sample region is performed, as illustrated in Fig. 18b. The white dotted boxes in Fig. 18a and Fig. 18b are respectively expanded in Fig. 18c and Fig. 18d. The former, Fig. 18c, reveals the enhanced contrast and super-resolution provided by EPSLON in comparison to the DL method. The latter, Fig. 18d, offers a complementary view of the same region, where the white arrows indicate features of fine correlation between EPSLON and SEM images. The mean FRC of DL image is 496 nm and of EPSLON is 176 nm, ∼2.8 times resolution gain. A Supplementary Movie 2 is provided to compare between EPSLON and SEM images in Fig. 18. One of the inherent advantages of photonic chip-based microscopy is large field-of-view (FOV) super-resolution imaging. Therefore, a larger FOV image of a human placenta section acquired using a 10X/0.25 NA and its corresponding EPSLON image, which is essential for histopathology applications, is shown in Supplementary Fig. 19.

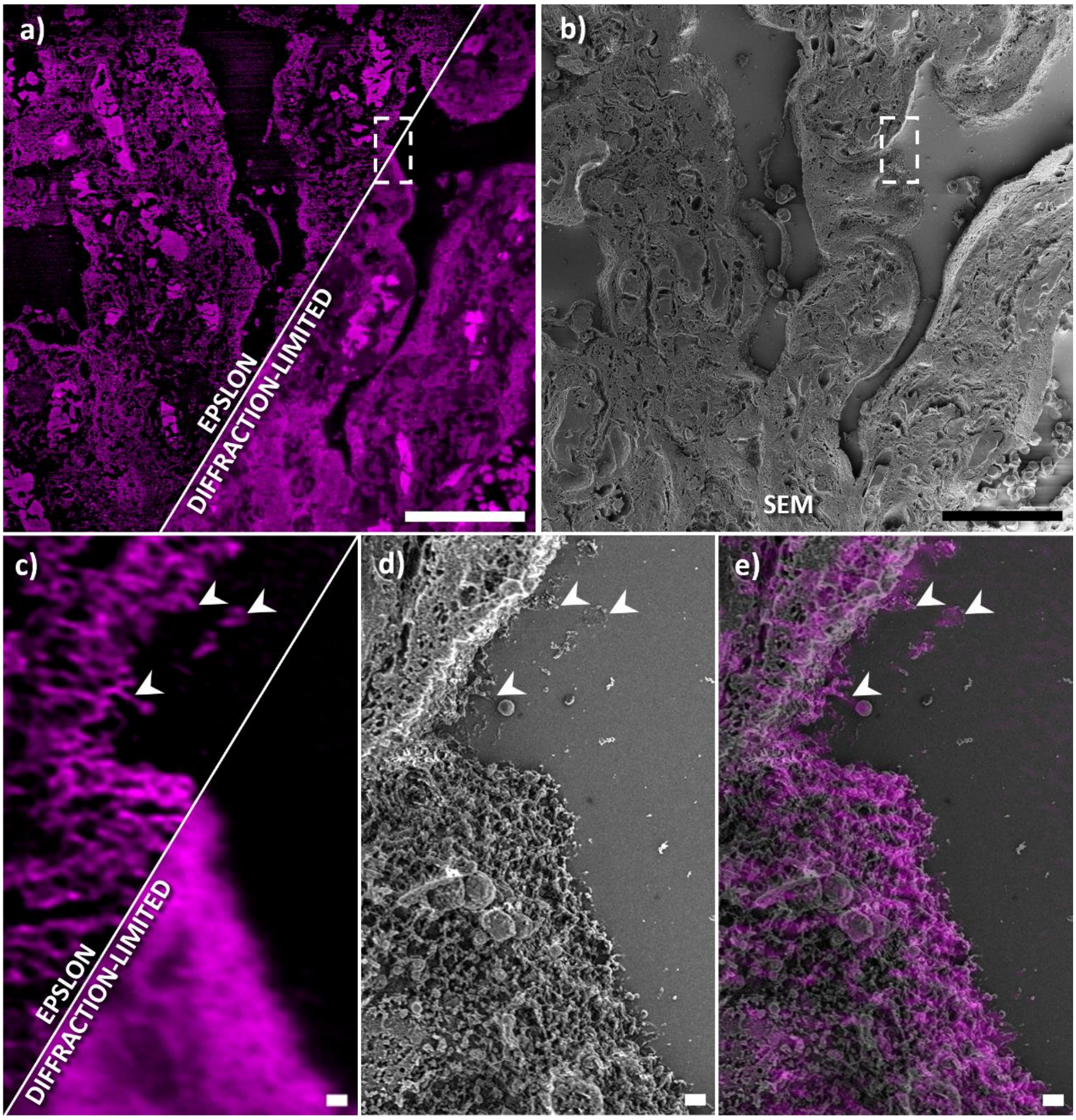

**Figure 19: EPSLON for label-free super-resolution imaging of human placenta tissue sections and benchmarking with correlative microscopy: EPLSON-SEM. (a)** A large FOV visualization of a human placental tissue section on a $Si_3N_4$ chip and imaged using a MO with detection NA = 1.2. Both DL and its corresponding super-resolved EPSLON images are shown, exhibiting a mean FRC resolution of 496 nm and 176 nm, respectively. **(b)** Same sample region acquired in a scanning electron microscope (SEM). **(c)** A zoomed-in view of the white-dotted box in (a) reveals the enhanced contrast and super-resolution provided by EPSLON in comparison to its DL counterpart. The white arrow heads denote the location of ultrastructural features with a high correlation with the SEM method. **(d)** Complementary SEM view of the same region in (c), illustrating with white arrow heads the correspondence with those seen via EPSLON. **(e)** An overlay view of (c) and (d) allows for visualizing correlation between EPSLON and SEM imaging methods. Scale bars a-b 50 µm, c-e 1 µm.

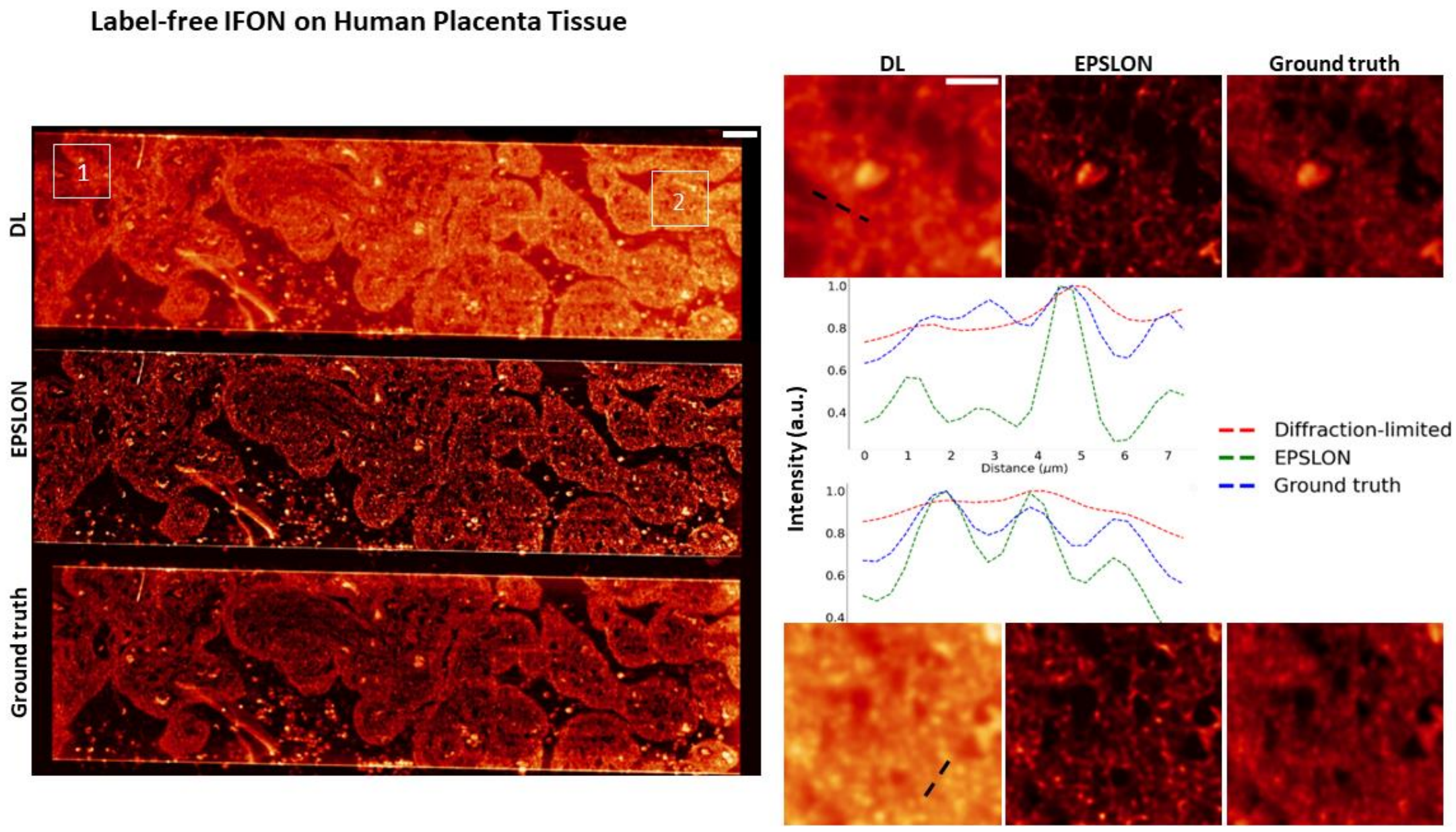

**Figure. 20: EPSLON for label-free super-resolution imaging of human placenta tissue sections**. Large field-of-view label-free diffraction-limited image termed DL, super-resolved EPSLON and ground truth images are shown, scale bar 25 µm. Two regions marked ‘‘1’’ and ‘‘2’’ in the DL image are blown-up and shown alongside. The corresponding regions in the EPSLON and ground truth images are also magnified and shown, scale bar 10 µm. Line profiles along the white dotted lines in the magnified boxes of the DL image fail to resolve any intricate features as shown by the line plots. EPSLON images provide more details as seen in the images and they are validated by the ground truth images, which are also evidenced by the line plots.

## 14. Experimental details

The following spectral filters are used in this work:

**Table 2:**

| Sl. No. | Filter Name | Spectral range (LP: LongPass, BP: BandPass) |
|---|---|---|
| 1 | FITC | 515 nm LP + 535\30 nm BP |
| 2 | TRITC | 585 nm LP \ 630\75 nm BP |
| 3 | CY5 | 655 nm LP \ 690\50 nm BP |

**Table 3:**

| Figure # | Sample | $\lambda_{ill}$ | Spectral filter \ $MO_2$ | Exposure | Reconstruction Algorithm | # Images | Comments |
|---|---|---|---|---|---|---|---|
| 3a | 195 nm polystyrene beads | 561 nm | TRITC \ 0.75 NA | 100 ms | BlindSIM[7] | 50 | |
| 3b | 100 nm Polystyrene beads | 405 nm | FITC \ 0.9 NA | 100 ms | SACD [8] | 100 | Order 2 used for reconstruction |
| 4 | (75 –250) nm Extracellular vesicles | 488 nm for PL | TRITC \ 0.45 NA | 1 sec | SACD | 50 | Order 2 used for reconstruction |
| | | 640 nm for TIRF | CY5 \ 0.45 NA, 1.2 NA | 100 ms | | 50 | |
| 5 | Rat kidney sections | 488 nm | FITC\ 1.42 NA | 100 ms | SACD | 250 | Order 2 used for reconstruction. |
| 6 | Human kidney section | 488 nm | FITC/0.9 NA | 100 ms | SACD | 250 | Order 2 used for reconstruction |
| Supplementary Fig. 10 | 195 nm polystyrene beads | 561 nm | TRITC \ 0.9 NA | 100 ms | BlindSIM | 50 | |
| Supplementary Fig. 11 | 200 nm gold nanoparticles | 640 nm | CY5\ 1.2 NA | 100 ms | FairSIM[9] | 3 | |
| Supplementary Fig. 19 | Human placenta tissue | 488 nm | TRITC \ 1.2 NA | 100 ms | SACD | 250 | Order 2 used for reconstruction |
| Supplementary Fig. 20 | Human placenta tissue | 488 nm | TRITC\ 0.25 NA (PL image) and 0.45 NA (Ground truth) | 50 ms | SACD | 20 | Order 2 used for reconstruction. Each image is acquired for 50 ms and 20 images are averaged to create (50 ms*20 = 1sec) to create one input image for SACD. |
| | | | | | | | |

## 15.References